\documentclass[12pt, a4paper, oneside]{article}
\usepackage[utf8]{inputenc}
\usepackage{amsmath}
\usepackage{amssymb}
\usepackage{graphicx}

\usepackage{float}
\usepackage{pdfpages}
\usepackage{caption}
\usepackage{subcaption}

\usepackage{xfrac}
\usepackage{hyperref}
\usepackage{xcolor}
\hypersetup{
    colorlinks,
    linkcolor={blue!50!black},
    citecolor={blue!50!black},
    urlcolor={blue!80!black}
}
\usepackage{amsthm}
\usepackage{algorithm}
\usepackage{algorithmicx}
\usepackage{algcompatible}
\usepackage{algpseudocode}
\usepackage{varwidth}
\usepackage{breqn}

\DeclareMathOperator*{\argmax}{arg\,max}

\usepackage{booktabs,makecell}
    
\usepackage{siunitx, mhchem}

\usepackage{tikz}
\usetikzlibrary{shapes,arrows}
\tikzstyle{block} = [rectangle, draw,
    text width=5em, text centered, rounded corners, minimum height=4em]
\tikzstyle{line} = [draw, -latex']    
\tikzstyle{optional}=[dashed,fill=gray!50]
\usepackage{multirow}

\usepackage{dsfont}
\usepackage[style=bwl-FU,maxbibnames=9,maxcitenames=1,uniquelist=false,
    backend=bibtex]{biblatex}
\usepackage{xpatch}
\usepackage{textpos}

\bibliography{ESGM-CSREM}

\usepackage{parskip}

\usepackage{amsmath,bm}
\usepackage{mathtools}
\usepackage{dirtytalk}
\usepackage{tikz}
\usetikzlibrary{shapes,arrows}
\tikzstyle{block} = [rectangle, draw,
    text width=3em, text centered, rounded corners, minimum height=2em]
\tikzstyle{line} = [draw, -latex']    
\tikzstyle{optional}=[dashed,fill=gray!50]
\tikzstyle{VineNode} = [ellipse, fill = white, draw = black, text = black, align = center, minimum height = 0.25cm, minimum width = 0.25cm]
\newcommand{\xshiftNodes}{0.5*\linewidth}
\newcommand{\yshiftLabels}{-.25cm}  
\newcommand{\labelsize}{\footnotesize} 

\usetikzlibrary{arrows.meta,chains,positioning,shapes.geometric}
\usepackage{float}

\usepackage{titling}
\usepackage{blindtext}
\usepackage{authblk}

\newtheorem{example}{Example}%
\newtheorem{definition}{Definition}%
\usepackage{titling}
\newcommand{\subtitle}[1]{%
  \posttitle{%
    \par\end{center}
    \begin{center}\large#1\end{center}
    \vskip0.5em}%
}

\begin{document}
\date{}
\title{Environmental, Social, Governance scores and the Missing pillar - Why does missing information matter?}
\subtitle{(ESGM: ESG scores and the Missing pillar)}
\author[1]{Özge Sahin$^{*,}$}
\author[2]{Karoline Bax}
\author[1,3]{Claudia Czado}
\author[2]{Sandra Paterlini}
\affil[1]{Department of Mathematics, Technical University of Munich, Boltzmanstraße 3, Garching, 85748, Germany}
\affil[2]{Department of Economics and Management, University of Trento, Via Inama 5, Trento, 38122, Italy}
\affil[3]{Munich Data Science Institute, Walther-von-Dyck-Straße 10, Garching, 85748, Germany}
\affil[ ]{\textit{*Corresponding author: ozge.sahin@tum.de}}

    \maketitle
    
    \begin{textblock*}{5cm}(0cm,-13cm)
  \fbox{\footnotesize To appear in  \textit{Corporate Social Responsibility and Environmental Management}}
    \fbox{\footnotesize DOI: \url{doi.org/10.1002/csr.2326}}
\end{textblock*}
    \begin{abstract}
Environmental, Social, and Governance (ESG) scores measure companies' performance concerning sustainability and are organized on three pillars: Environmental, Social, and Governance. These complementary non-financial ESG scores should provide information about companies' ESG performance and risks. However, the extent of not yet published ESG information makes the reliability of ESG scores questionable. To explicitly capture the not yet published information on ESG category scores, a new pillar, the so-called Missing (M) pillar, is proposed and added to the new definition of the Environmental, Social, Governance, and Missing (ESGM) scores. By relying on the data provided by Refinitiv, we show that the ESGM scores strengthen the companies' risk relationship. These new scores could benefit investors and practitioners as ESG exclusion strategies using only ESG scores might exclude assets with a low score solely because of their missing information and not necessarily because of a low ESG merit.
    \end{abstract}

\newpage
\section{Introduction}\label{intro}
As sustainability concerns increase globally, sustainable finance and Environmental, Social, and Governance (ESG) investing strategies gained much interest. According to Bloomberg, the \say{ESG ETF market had risen over 318\% in 2020}, indicating the significant interest by investors \parencite{Bloomberg}. To assess the companies' ESG performance and sustainability, investors can use the ESG scores data providers make available. Such scores use the publicly available data and voluntary disclosure to compute individual \textit{Environmental (E)}, \textit{Social (S)}, and \textit{Governance (G)} pillar scores as well as an overall ESG aggregated score. 

However, in the last years, criticism of ESG scores has emerged. These include a large discrepancy between ESG scores from different data providers (e.g., \cite{Berg2019, Billio2021, Gyonyorova2021}), as well as the possible update of ESG scores within five years from the data provider (e.g., Thomson Reuters (Refinitiv) scores are only definitive after five years). Moreover, ESG scores might be subject to changes due to a release of new ESG information, i.e., the release of missing ESG information (e.g., \cite{Berg2021, Sahin2022}). 

In this paper, we focus on studying the role and the amount of missing ESG information as a potential source for a release of new ESG information with impacts on ESG scores in the future.  Thus, we introduce a new pillar, called the \textit{Missing (M) pillar}, and define new scores: \textit{Environmental, Social, Governance, Missing (ESGM)} scores by simultaneously aggregating the M pillar with the three ESG pillars. The ESGM scores are easily interpretable as a convex combination of the E, S, G, and the newly introduced M pillar scores. We propose an optimization scheme to link ESGM scores and risk measures and run an in-sample and out-of-sample analysis to ensure robust results. If the amount of missing information is explicitly considered, companies are encouraged to disclose new information by our ESGM score construction methodology as it positively impacts the score. This assumption is reasonable considering the current ESG score construction methodology that positively rewards the disclosure of new information and the fact that often missing information is not due to the unwillingness to release such information but its unavailability. 

Refinitiv is a key data provider whose ESG scores are used by many scholars and investors (e.g., \cite{Berg2021}). We work with the Refinitiv ESG data of the constituents of the S\&P 500 and EuroStoxx 600 in the period 2017-2019, i.e., when the missing ESG information can still be released, and the ESG scores can be updated.  We show that ESG and risk dependence and the amount of the missing ESG information change with sectors and geographical regions. We also show that ESGM scores provide better risk profiles for companies than ESG scores.

Moreover, investors and practitioners can benefit from this research, as negative screening as an investment strategy (only including companies with a high ESG score in a portfolio and excluding companies with a low ESG score (e.g., \cite{Pri}) is well established. However, following this approach and using the widely available Refinitiv ESG data, which do not include the potential of new information disclosure, would mean that possible companies with potentially high scores after new ESG information adoption will be missed. Overall, this could lead to a more risky and less effective portfolio. ESGM scores identify the risky companies better than ESG scores for exclusion strategies. Nonetheless, the impact of missing ESG information on the negative screening varies in sectors and regions. We further discuss the implications of our research for researchers, investors, companies, and managers.

The paper is structured as follows. We first review the related literature in Section \ref{litrew} and then introduce the methodology behind ESG score construction from Refinitiv in Section \ref{refinitiv}. Section \ref{method} describes our methodology for \textit{M Pillar} and ESGM scores, while Section \ref{results} reports our empirical results. Section \ref{conc} concludes.  

\section{Literature review}\label{litrew}
Until now, most scholars have focused on the link between ESG scores and corporate financial performance (e.g., \cite{Friede2015, Cornett2016,  Henke2016, Ghoul2017, Hou2019, Behl2021, Kalaitzoglou2021}). Then recently, some studies have started to analyze  the link between ESG scores and risk measures (e.g., \cite{Shafer2020, Bax2021, Giese2021, Maiti2021}).  Additionally, ESG data quality issues and the impact of ESG-type corporate disclosures on investment allocations have been another center of attention in the ESG literature and which is where our work contributes.

\cite{Berg2019} report a large discrepancy between ESG scores from different data providers. \cite{Abhayawansa2021} present the main reason for the divergence as different measurement methods. \cite{Gyonyorova2021} discuss that such a divergence changes from sector to country. Finally, \cite{Billio2021} argue that such discrepancies might make the usage of ESG scores in portfolios difficult for fund managers.  However, not only the critique towards different providers has been rising, but also has the critique towards single data providers. Recently, \cite{Berg2021} noticed changes in the historical ESG scores given by Refinitiv. Additionally, \cite{Sahin2022} discuss how the scores might differ for the five most recent years. Such changes, for instance, can arise from the release of missing information.

While the above-described debate is still going on,  ESG scores still play a crucial role in investors' investment strategies. Based on the global survey conducted by senior investment professionals, \cite{Amel2018} find that the negative screening, either across sectors or within a sector, is  still the most used method to integrate ESG information into portfolios compared to positive screening, active ownership, and full integration.  As an outlook on the future, investors argue that they expect positive screening and active ownership to gain importance \parencite{Amel2018}. Moreover, \cite{Alessandrini2020} and \cite{Alessandrini2021} discuss that the performance of the ESG exclusion strategies varies across geographies and sectors.  They find that screening often leads to a better risk profile of the portfolios and often generates protection against credit risks \parencite{Alessandrini2021}. Lastly, they recommend screening as the best strategy for passive investors with ESG preferences \parencite{Alessandrini2021}. Still, the debate about the negative screening has been ongoing since the exclusion strategies based on ESG scores can lead to capital and risk misallocations if ESG scores are not representative of company characteristics \parencite{Alessandrini2020}.

\section{Review of Refinitiv's ESG score methodology}\label{refinitiv}
Refinitiv collects publicly available ESG information of companies and aggregates such information to assign the companies with ten ESG category scores benchmarked against Thomson Reuters Business Classifications Industry Group or the respective Country Group \parencite{Refinitiv2021, Refinitiv2021b}. The ten categories are  Environmental Innovation (EI), Resource Use (RU), Emissions (EM), Workforce (WF), Human Rights (HR), Community (CO),  Product Responsibility (PR), Management (MG), Shareholders (SH), and Corporate Social Responsibility (CS). 

Then, the ESG category scores are aggregated to build the \textit{Environmental (E)}, \textit{Social (S)}, and \textit{Governance (G)} pillar scores as illustrated in Figure 1. As a result, each pillar score is between zero and 100.

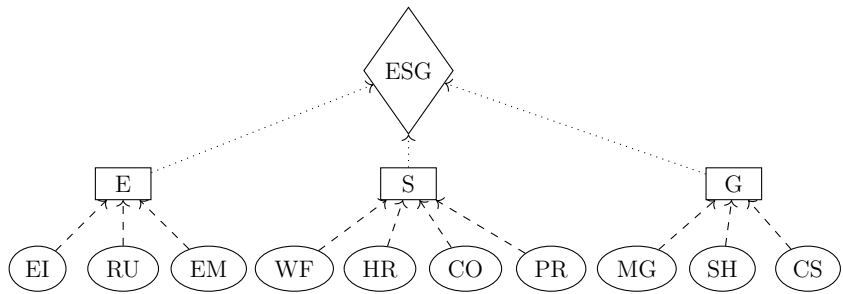
\begin{figure}[H]
         \caption{Aggregation of ten ESG categories and three pillar scores in Refinitiv's ESG score construction methodology.}
	\centering
	\renewcommand{\xshiftNodes}{0.0001*\linewidth}
	\renewcommand{\yshiftLabels}{.0cm}  
	\renewcommand{\labelsize}{\scriptsize} 
  \begin{tikzpicture}[node distance =1.5cm,scale=0.75, transform shape,
  disc/.style = {shape=diamond, draw, shape aspect=0.7,shape border rotate=90,
  text width=9mm, align=center},
  alg/.style = {shape=rectangle, draw, shape aspect=0.3,shape border rotate=90,
  text width=7mm, align=center}]
\node (v1) [VineNode] {EI}
node (v2) [VineNode, right of = v1, xshift = \xshiftNodes]  {RU}
node (v3) [VineNode, right of = v2, xshift = \xshiftNodes]  {EM}
node (v4) [VineNode, right of = v3, xshift = \xshiftNodes] {WF}
node (v5) [VineNode, right of = v4, xshift = \xshiftNodes] {HR}
node (v6) [VineNode, right of = v5, xshift = \xshiftNodes] {CO}
node (v7) [VineNode, right of = v6, xshift = \xshiftNodes] {PR}
node (v8) [VineNode, right of = v7, xshift = \xshiftNodes] {MG}
node (v9) [VineNode, right of = v8, xshift = \xshiftNodes] {SH}
node (v10) [VineNode, right of = v9, xshift = \xshiftNodes] {CS}
node (v11) [alg, above of = v2, yshift= \xshiftNodes] {E}
node (v12) [alg, above of = v5, yshift = \xshiftNodes,  xshift =  0.5cm] {S}
node (v13) [alg, above of = v9, yshift =\xshiftNodes, xshift=0.2cm] {G}
node (v15) [disc, above of = v12, yshift=0.5cm] {ESG};
\draw[dashed,->, color=black] (v1) to node[draw=none,  font = \labelsize,fill = none, below, yshift = \yshiftLabels] {} (v11);
\draw[dashed,->, color=black] (v2) to node[draw=none,  font = \labelsize,fill = none, below, yshift = \yshiftLabels] {} (v11);
\draw[dashed,->, color=black] (v3) to node[draw=none,  font = \labelsize,fill = none, below, yshift = \yshiftLabels] {} (v11);
\draw[dashed,->, color=black] (v4) to node[draw=none,  font = \labelsize,fill = none, below, yshift = \yshiftLabels] {} (v12);
\draw[dashed,->, color=black] (v5) to node[draw=none,  font = \labelsize,fill = none, below, yshift = \yshiftLabels] {} (v12);
\draw[dashed,->, color=black] (v6) to node[draw=none,  font = \labelsize,fill = none, below, yshift = \yshiftLabels] {} (v12);
\draw[dashed,->, color=black] (v7) to node[draw=none,  font = \labelsize,fill = none, below, yshift = \yshiftLabels] {} (v12);
\draw[dashed,->, color=black] (v8) to node[draw=none,  font = \labelsize,fill = none, below, yshift = \yshiftLabels] {} (v13);
\draw[dashed,->, color=black](v9) to node[draw=none,  font = \labelsize,fill = none, below, yshift = \yshiftLabels] {} (v13);
\draw[dashed,->, color=black](v10) to node[draw=none,  font = \labelsize,fill = none, below, yshift = \yshiftLabels] {} (v13);
\draw[dotted, ->,color=black] (v11) to node[draw=none,  font = \labelsize,fill = none, below, yshift = \yshiftLabels] {} (v15);
\draw[dotted, ->, color=black] (v12) to node[draw=none,  font = \labelsize,fill = none, below, yshift = \yshiftLabels] {} (v15);
\draw[dotted, ->, color=black] (v13) to node[draw=none,  font = \labelsize,fill = none, below, yshift = \yshiftLabels] {} (v15);
    \end{tikzpicture}
    \label{fig:esg-tikz}
\end{figure}

Next, an overall ESG aggregated score is the weighted sum of three pillar scores, i.e., a convex combination of the E, S, and G pillar scores. The pillar score weights range from zero to one, sum up to one, and can change for each pillar within industry groups (see Pages 9 and 10 in \cite{Refinitiv2021}). Hence, overall ESG aggregated scores also range from zero to 100. The higher the overall ESG aggregated scores are, the more ESG responsible the company is evaluated. In Example \ref{ex:ESG}, we present an exemplary overall ESG aggregated score calculation of Refinitiv using the fictitious pillar score weights of the industry group Household Goods and a generic name for the company. 

\begin{example} (Overall ESG aggregated score calculation) 
The ESG pillar scores of Company F in Household Goods in 2017 are 0.00 (E pillar), 63.01 (S pillar), and 54.77 (G pillar) with weights 0.240, 0.294, and 0.466, respectively. Then its overall ESG aggregated score in 2017 is the weighted sum of all pillar scores: \\
$
x_{ESG, CompanyF, 2017} = 0.240 \cdot 0.00 + 0.294 \cdot 63.01 + 0.466 \cdot 54.77  = 44.05.
$
\label{ex:ESG}
\end{example}
\newpage
A drawback of the current ESG score methodology is that it does not explicitly consider the potential disclosure of new ESG information. Indeed, Company F in Example \ref{ex:ESG} has not yet provided any ESG information for the E pillar score of 2017; therefore, its E pillar score is zero. It also implies that the ESG category scores of 2017 building the E pillar score (Resource Use, Emissions, and Environmental Innovation) are zero. Company F can disclose its ESG information regarding the E pillar's categories until June 2022, given that ESG data for the last five years can be updated a posteriori \parencite{Refinitiv2021}. 

To quantify the companies' potential to disclose more ESG information in time and analyze the impact of the disclosure on the ESG scores and the risk of the portfolios, we construct our methodology in Section \ref{method}. Overall ESG aggregated scores will be called ESG scores in the rest of the paper.

\section{Research methodology}\label{method}
In the following section, we formulate a \textit{Missing (M-) pillar} score, which explicitly captures the amount of not yet reported ESG information regarding the ten ESG categories. Later, we define \textit{Environmental, Social, Governance, and Missing (ESGM)} scores and propose an optimization approach for their computation, linking them to companies' riskiness. The optimization scheme aims to harvest the potential to strengthen the risk relationship in disclosing missing ESG information \parencite{EBA2021}. We report the notations of data and indices used in the paper in Appendix \ref{app-notation}. 

\subsection{Missing pillar (M pillar) score}\label{m-pillar}
Since a zero ESG category score denotes that the company has not yet reported any information regarding it as discussed in Example \ref{ex:ESG}, we define a new pillar accounting for zero values, i.e., missing information, in the ten ESG category scores in a given business class: the \textit{Missing (M-) pillar} score. A business class can be an industry group or an economic sector based on the business classification of Thomson Reuters \parencite{Refinitiv2021}. For instance, the assets selected from S\&P 500 in Section \ref{data} belongs to 61 industry groups from ten economic sectors.
\begin{definition} {(The M pillar score)} 
For company $p$ in a given business class $a$ and year $t$, first, we find the total number of zero values in its ten ESG categories, i.e., $x^{a}_{zero,p, t}$. The set $S^{a}_{zero, t}$ contains these values for all $a, p,t$. Denoting the total number of companies with same value as $p$ in $S^{a}_{zero, t}$ (including the company $p$ itself) by $e^{a}_{p,t}$, and the total number of companies with a higher value than $p$ in $S^{a}_{zero, t}$ by $l^{a}_{p,t}$, the M pillar score of company $p$ in a given business class $a$ and year $t$ is defined as:
\begin{equation}
x^{a}_{M, p, t} = 100 \cdot \frac{l^{a}_{p,t} + \frac{e^{a}_{p,t}}{2}}{n_a}, \quad p =1, \ldots, n_a, \quad  t=1,\ldots, T, \quad  a=1, \ldots, A,
\label{eq:m-pillar}
\end{equation}
where $n_a$ is the total number of companies in the given business class $a$, $T$ is the total number of years, and $A$ is the total number of given business classes. The detailed calculations and notations are given in Appendix \ref{app-notation}. 
\end{definition}

From the definition, when a company has more zero values in its ESG categories than all other companies in its business class, its M pillar score will be the highest. This is because it has a higher extent of not yet published ESG information reflected in a high M pillar score. Moreover, the M pillar score is continuous and between zero and 100, with a mean value of 50 (proof in Appendix \ref{app-proof}). Such a formulation makes it robust to outliers and comparable with the three ESG pillar scores. Its formulation is similar to our data provider's ESG category score methodology \parencite{Refinitiv2021}. 

Since our data provider has ten ESG category scores, the highest total number of zero values a company can have in its ESG categories in our empirical analysis in Section \ref{results} is ten. Accordingly, Example \ref{ex:m-pillar} shows the M pillar score calculation steps using ten ESG category scores, where the business class is an economic sector, e.g., Consumer Cyclicals.  

\begin{example} (The M pillar score calculation) 
Let $\bm{x}^{a}_{p, t} =(x^{a}_{RU, p,t}, \ldots, x^{a}_{CS, p,t})^\top$ contain ten ESG category scores of company $p$ in business class $a$ in year $t$. Suppose there are four companies ($n_a=4$) in the economic sector Consumer Cyclicals ($a=2$), and their fictitious ESG category scores in 2017 ($t=2017$) are given as follows: \\
 $\bm{x}^{2}_{1, 2017} = (99.3 , 50.1 , 12.3 , 52.2 , 0.00, 67.9 , 0.00 , 11.2 , 20.4 , 0.00)^\top$,  \\
 $\bm{x}^{2}_{2, 2017} = (63.5 , 70.1 , 52.3 , 84.3 , 10.2, 77.9 , 88.9 , 55.2 , 80.4 , 86.3)^\top$, \\
 $\bm{x}^{2}_{3, 2017} = (36.3 ,0.00 , 12.3 , 23.2 , 0.00, 17.9 , 0.00 , 21.2 , 50.5 , 58.3)^\top$, \\
 $\bm{x}^{2}_{4, 2017} = (85.2 , 0.00, 12.3 , 12.2 , 0.00, 54.3 , 52.5 , 81.2 , 75.6 , 24.3)^\top$. \\
For company $p$ with $p=1, \dots, 4$, we determine the total number of zero values in its ESG categories and have $S^{2}_{zero, 2017}= \{3, 0,3,2\}$. Consider the fourth company: it holds $x^{2}_{zero,4, 2017}=2$, $e^{2}_{4, 2017} =1$, and $l^{2}_{4, 2017} =1$. Accordingly, we calculate its M pillar score as given in Equation \eqref{eq:m-pillar}:  $x^{2}_{M, 4, 2017} = 100\cdot\frac{1 + \frac{1}{2}}{4} =37.5$. For the second company without any zero values in its ESG categories, the M pillar score is given by $x^{2}_{M, 2, 2017} = 100\cdot\frac{0 + \frac{1}{2}}{4} = 12.5$. Since we also calculate $x^{2}_{M, 1, 2017} = 75$ and $x^{2}_{M, 3, 2017} =75$, the M pillar score is between zero and 100; the average M pillar score of all companies is 50, as postulated.
\label{ex:m-pillar}
\end{example}

\subsection{Environmental, Social, Governance, and Missing (ESGM) scores} \label{esgm}
This section incorporates the M pillar into the three ESG pillars and builds new scores, the \textit{ESGM: Environmental, Social, Governance, and Missing} scores. 

\begin{definition} {(The ESGM score)} 
The ESGM score of company $p$ in a given business class $a$ and year $t$ is defined as a weighted sum:
	\begin{equation}
x^{a}_{ESGM, p, t} = x^{a}_{E, p, t} \cdot w^{a}_{E} + x^{a}_{S, p, t} \cdot w^{a}_{S} + x^{a}_{G, p, t} \cdot w^{a}_{G} + x^{a}_{M, p, t} \cdot w^{a}_{M}, \quad \forall{a,p,t}.
\label{eq:esgm}	
	\end{equation}
\end{definition}

The ESGM scores have four weighted pillar scores, and the unknown pillar score weight varies according to its business class. The next task is to estimate the pillar score weights. Since even regulatory authorities, including the European Banking Authority (EBA), have acknowledged the role of ESG scores in quantifying the company's riskiness and have identified a need to incorporate ESG risks into overall business strategies and risk management frameworks \parencite{EBA2021}, we propose the following optimization scheme in Equation \eqref{eq:opt-problem-sec-esg} to estimate them, connecting the companies' ESGM scores with their risk performance. 

Refinitiv allows investors to build custom overall aggregated ESG scores by assigning customized pillar weights \parencite{Refinitiv2021c}. Therefore, our proposed optimization scheme that links the scores and riskiness can be applied for such a custom aggregation by investors. From the angle of corporate investments' net present value estimation process, \cite{Kudratova2020} also presented an optimization model for quantitative sustainability measurements. 

\begin{subequations}
\begin{alignat}{2}
(\hat{w}^{a}_{E}, \hat{w}^{a}_{S},\hat{w}^{a}_{G}, \hat{w}^{a}_{M}) = &\!\argmax_{\substack{w^{a}_{E}, w^{a}_{S}, w^{a}_{G}, w^{a}_{M} }}        &\qquad &\sum_{t=t_{1}}^{t_{2}} \hat{\tau}_{risk}\Big(  \bm{x}^{a}_{ESGM, t}, \quad \bm{x}^{a}_{risk, t}\Big)\label{eq:optProb}\\
&\text{subject to}&&w^{a}_{E} + w^{a}_{S}+ w^{a}_{G}+ w^{a}_{M}= 1,\quad \forall{a},\label{eq:constraint0}\\
&&&w^{a}_{E} , w^{a}_{S}, w^{a}_{G}  \geq 0.100, \quad \forall{a},\label{eq:constraint2}\\
& && w^{a}_{M}  \geq 0, \quad \forall{a},\label{eq:constraint1} \\ 
&&&w^{a}_{E} , w^{a}_{S}, w^{a}_{G}  \geq w^{a}_{M}, \quad \forall{a}.\label{eq:constraint3}
\end{alignat}
\label{eq:opt-problem-sec-esg}
\end{subequations}

We analyze the ESGM scores’ influence on their risk performance by focusing on their dependence in Equation \eqref{eq:optProb}. We choose Kendall's tau ($\tau$) as our dependence measure since it is robust to outliers. Given the growing literature linking the ESG scores with a risk measure, such as the VaR (e.g., \cite{Verheyden2016}) or volatility (e.g., \cite{Zhang2021}), we can specify a generic risk function in Equation \eqref{eq:risk-func}. More complex objective functions aiming to find optimal ESG portfolios for investors, such as proposed in \cite{Ahmed2021} and \cite{Pedersen2021}, are subject to future research. Nevertheless, we propose a flexible framework that can take only VaR and volatility or their joint interaction with the artificially introduced risk measure as the product of the VaR and volatility, i.e., \textit{vvrisk}. High ESGM scores should be linked to the strong VaR (e.g., \cite{Diemont2016}) and \textit{vvrisk}, as well as the low volatility (e.g., \cite{Kumar2016}). Moreover, the pillar score weights can be estimated using data from the period $[t_1, t_2]$. In Section \ref{in-sample}, we will be using the two recent years of the ESG data for an in-sample estimation, i.e., $t_1= 2017,  t_2= 2018$, while $t_3=2019$ will be used for an out-of-sample evaluation in Section \ref{out-of-sample}.

\begin{equation}
\hat{\tau}_{risk}\Big(  \bm{x}^{a}_{ESGM, t}, \quad \bm{x}^{a}_{risk,t}\Big) = 
\begin{cases} 
         & \hat{\tau}\Big(  \bm{x}^{a}_{ESGM, t}, \quad \bm{x}^{a}_{vv,t}\Big),  \quad    \quad \textrm{if } risk = vvrisk, \\
       &  \hat{\tau}\Big(  \bm{x}^{a}_{ESGM, t}, \quad \bm{x}^{a}_{VaR,t}\Big),  \quad  \textrm{if } risk = VaR, \\
       & -\hat{\tau}\Big(  \bm{x}^{a}_{ESGM, t}, \quad \bm{x}^{a}_{vol,t}\Big), \quad  \textrm{if } risk = vol.
\end{cases} 
\label{eq:risk-func}
\end{equation}

The constraint in Equation \eqref{eq:constraint0} ensures that the ESGM scores are between zero and 100, similar to the ESG scores. To exclude unrealistic scenarios, Equation \eqref{eq:constraint2} ensures that each pillar score except the M pillar score has a positive lower bound, which is motivated by the lowest weight ever given to one of the E, S, and G pillar scores in one of the industry groups by our data provider. To account for cases with no impacts of disclosing new ESG information on the risk performance, we set a lower bound of zero for the M pillar score weight in Equation \eqref{eq:constraint1}. In Equation \eqref{eq:constraint3}, we assume that the E, S, and G pillar score weights are larger than or equal to the M pillar score weight. It considers the relative importance of already disclosed ESG information in the E, S, and G pillar scores compared to the potential disclosure represented by the M pillar score. 

Moreover, disclosing reduces the companies' weighted M pillar score due to a decrease in the number of zero values entering the computation of the M pillar score. By assigning higher weights to the E, S, and G pillar scores, our scheme encourages companies to disclose new ESG information, which usually positively impacts both their ESGM and ESG scores. Additionally, such an optimization scheme allows us to see which business classes with not yet disclosed ESG information might play a role and for which business classes a re-weighting scheme for the E, S, and G pillar scores matters to strengthen the risk dependence. An empirical analysis is provided in Section \ref{results}.

Overall, in Equation \eqref{eq:opt-problem-sec-esg}, the constraints are linear, and the objective function is non-linear in terms of the parameters with unknown derivatives. Thus, such a scheme can be solved by a derivative-free optimization algorithm dealing with linear constraints. \cite{Larson2019} provide a recent review of derivative-free optimization methods.

When our ESGM pillar score weights are estimated,  we compute a company's ESGM score as follows in Example \ref{ex:esgm}.

\begin{example} (The ESGM score calculation) \\
Suppose the estimated pillar weights of the companies in the economic sector Consumer Cyclicals ($a=2$) are given by $w^{2}_{E} =0.258$, $w^{2}_{S} =0.122$, $w^{2}_{G} =0.498$, $w^{2}_{M} =0.122$. Then, a company's ESGM score ($p=1$) in Consumer Cyclicals with the following pillar scores in 2017,  $x^{a}_{E, 1, 2017}=40.0$, $x^{a}_{S, 1, 2017}=60.0$, $x^{a}_{G, 1, 2017}=20.0$, $x^{a}_{M, 1, 2017}=50.0$, is calculated as follows:\\
$x^{2}_{ESGM, 1, 2017} = 0.258 \cdot 40.0 + 0.122 \cdot 60.0 + 0.498 \cdot 20.0+ 0.122  \cdot 50.0 = 33.70$.
\label{ex:esgm}
\end{example}
\newpage
\section{Empirical results}\label{results}
Before reporting our empirical results, we first explore the data in this section. The first and second data sets consist of the constituents of the S\&P 500 and EuroStoxx 600 from ten economic sectors, i.e., the top market capitalization companies in the USA and Europe, respectively. For each data set, we calculate the M pillar score and estimate the ESGM pillar weights by solving the optimization scheme in Equation \eqref{eq:opt-problem-sec-esg}, using the derivative-free optimization solver, LINCOA (Linearly Constrained Optimization Algorithm) \footnote{LINCOA solves linearly constrained optimization problems without using derivatives of the objective function and uses a trust region method (\cite{Powell2015}). As \cite{Powell2015} mentioned, we transform the linear equality in Equation \eqref{eq:constraint0} into two inequalities. After running sensitivity analyses, the initial and final trust-region radii are set to $0.2$ and $0.0005$, respectively. The maximum number of function evaluations allowed is 10,000. As the numerical optimization problems are sensitive to initial parameter values, we use ten different starting values and choose the optimal weights in correspondence with the best objective function value of ten runs. We do not observe multiple optimal solutions. All results are available from the authors upon request.}. Later, we compute the ESGM scores in each data set for each of the ten sectors. We base our four pillar score weights estimation on training (in-sample) data to avoid overfitting the data. Finally, we compare the relationship between risk, ESGM scores and ESG scores using test (out-of-sample) data.

\subsection{Data}\label{data}
Using the non-definitive ESG data, for which companies can still disclose the ESG information, our data consists of yearly ESG, E, S, and G pillars composed of the ten ESG categories Resource Use, Emissions, Environmental Innovation, Workforce, Human Rights, Community, Product Responsibility, Management, Shareholders, and Corporate Social Responsibility scores of the constituents of the S\&P 500 (extracted on February 4, 2021) and the constituents of the EuroStoxx 600 (extracted on March 28, 2022) over the period 2017 to 2019. 

We use the companies' daily price data from January 2, 2017, to December 30, 2019, to compute their daily log returns. Since 17 companies in the S\&P 500 and 109 companies in the EuroStoxx 600 do not report either ESG data or price data in 2017-2018 in the database, we excluded them from our analysis, working with 483 companies in the S\&P 500 and 491 companies in the EuroStoxx 600. To have as many companies as possible in the sample, we argue that investors use the latest score available in the market to make their risk assessment. Hence, the ESG data of the companies, which do not have any values in 2019, is imputed by their ESG data in 2018, assuming their score has not yet been updated, and investors would still consider these scores in their decision making.  In total, we imputed the ESG data of 66 companies in the S\&P 500 and 19 companies in the EuroStoxx 600 in 2019. The dependence estimated using Pearson correlation (Kendall’s $\tau$) between the ESG scores in 2017 and 2018 is equal to 0.94 (0.79), between the scores in 2018 and 2019 to 0.93 (0.78), while between the ESG scores in 2017 and 2019 to 0.89 (0.71) in the S\&P 500. Similar results apply for the EuroStoxx 600.

Both samples include companies from ten different Thomson Reuters Business Classifications Economic Sectors (\cite{Refinitiv2021b}): Basic Materials (23 and 50 companies), Consumer Cyclicals (77 and 79 companies), Consumer Non-Cyclicals (39 and 41 companies), Energy (24 and 17 companies), Financials (60 and 88 companies), Healthcare (56 and 33 companies), Industrials (65 and 84 companies), Real Estate (28 and 26 companies), Technology (82 and 45 companies), and Utilities (29 and 28 companies) in the S\&P 500 and EuroStoxx 600, respectively. Even though the data provider determines the ESG, pillar, and category scores for 47 industry groups  within the S\&P 500 and 52 industry groups within the EuroStoxx 600 as the business class, we work with ten economic sectors to have a larger sample size within each sector to optimize the ESGM pillar score weights. Since using industry groups is a more granular approach than using sectors due to a larger number of convex pillar score weight combinations, it implies a trade-off in favor of the data provider's weighting scheme. Still, the ESGM scores show a risk strengthening effect in Sections \ref{in-sample} and \ref{out-of-sample} compared to the data provider's ESG scores. 

Table 1 shows that the percentage of the companies with not yet disclosed ESG information regarding at least one of the ten ESG categories ranges from a minimum of 15\% in Consumer Non-Cyclicals in 2019 to a maximum of 71\% in Healthcare in 2017, with an average of 47\% across ten sectors and three years in the S\&P 500. Its range is from 11\% in Utilities in 2019 to 77\% in Healthcare in 2017 with an average of 38\% across ten sectors and three years in the EuroStoxx 600. Additionally, it is always higher in the S\&P 500 than the EuroStoxx 600 per year in all sectors but Real Estate. More than half of the companies in Consumer Cyclicals, Energy, Healthcare  in the S\&P 500 and Real Estate in the EuroStoxx 600 have at least one of the ten ESG categories with not yet released ESG information each year. Moreover, we observe that the percentage of such companies tends to decrease in time in Table 1. Such a result might imply that the companies disclose more information as it becomes available, which could provide new insights into their ESG performance and risk characteristics. Furthermore, since the ESG scores have had a strong impact on the company's value (e.g., \cite{Fatemi2018}), one can expect the companies to publish more ESG information in the future. 

\begin{table}[H]
\centering
\begin{tabular}{lcccc}
\hline
Sector (S\&P) & 2017 & 2018 & 2019  \\ 
  \hline
   Basic Materials & 35\% & 30\% & 30\% \\ 
 Consumer Cycl. & 56\% & 55\% & 53\% \\ 
   Consumer N-Cycl. & 23\% & 18\% & 15\%  \\ 
     Energy &67\% & 67\% & 58\% \\ 
        Financials & 62\% & 53\% & 47\% \\ 
  Healthcare & 71\% & 64\% & 59\%\\ 
   Industrials & 48\% & 48\% & 45\%  \\ 
     Real Estate & 68\% & 54\% & 46\% \\ 
 Technology & 48\% & 44\% & 34\%  \\ 
  Utilities & 45\% & 41\% & 34\% \\ 
   \hline
\end{tabular}
\quad
\begin{tabular}{lcccc}
\hline
Sector (EuroStoxx) & 2017 & 2018 & 2019  \\ 
  \hline
 Basic Materials & 36\% & 24\% & 20\% \\ 
  Consumer Cycl. & 43\% & 35\% & 30\% \\ 
  Consumer N-Cycl. & 37\% & 34\% & 22\% \\ 
  Energy & 35\% & 35\% & 29\% \\ 
  Financials & 50\% & 36\% & 26\% \\ 
   Healthcare & 70\% & 58\% & 45\% \\ 
 Industrials & 46\% & 37\% & 30\% \\ 
   Real Estate & 77\% & 65\% & 54\% \\ 
Technology & 42\% & 36\% & 33\% \\ 
 Utilities & 32\% & 18\% & 11\% \\ 
   \hline
\end{tabular}
\caption{Percentage of the companies with not yet disclosed ESG information at least one of the ten ESG categories across ten sectors and three years in S\&P 500 (left) and EuroStoxx 600 (right).}
\label{table1}
\end{table}   

Figure 2 shows that companies with lower ESG scores tend to have more ESG categories with not yet published ESG information. It could be due to the lack of infrastructure allowing them to collect and then release such information. However, it would still imply that a company with a lower ESG score could have a large potential to upgrade its ESG score when not yet recorded information is disclosed.

\begin{figure}[H]
\centering
     \caption{Scatter plot of the companies' ESG scores and their number of ESG categories with undisclosed ESG information in Consumer Cyclicals in S\&P 500 in 2017, where a diamond denotes the median ESG score of the respective column.}
\includegraphics[width=.75\textwidth]{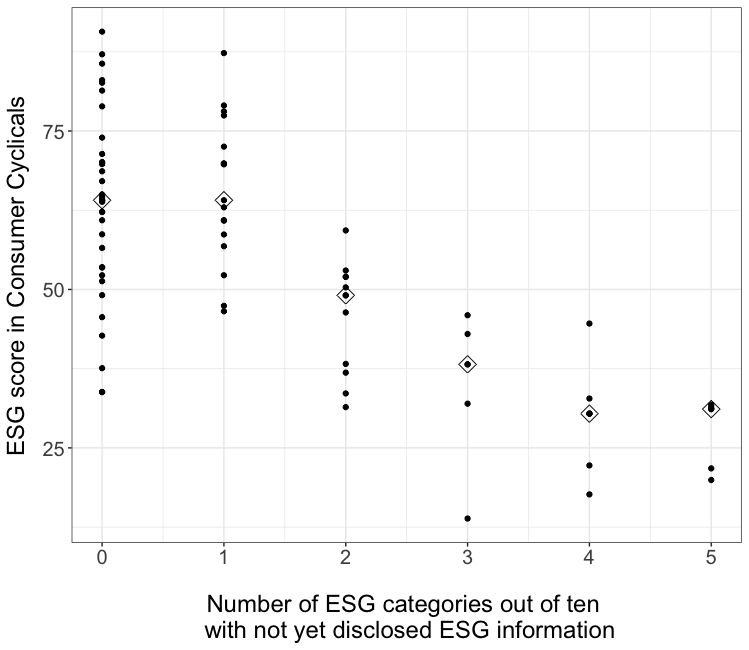}
   \label{fig:cc-miss}
\end{figure}

Figure 3 shows the variability in the companies' E, S, and G pillar scores in 2017 across ten sectors in the S\&P 500. We observe similar findings for 2018 and 2019 and the data set EuroStoxx 600 (available upon request).

\begin{figure}[H]
\centering
     \caption{E, S, and G pillar scores in S\&P 500 in 2017 across ten sectors.}
\includegraphics[width=\textwidth]{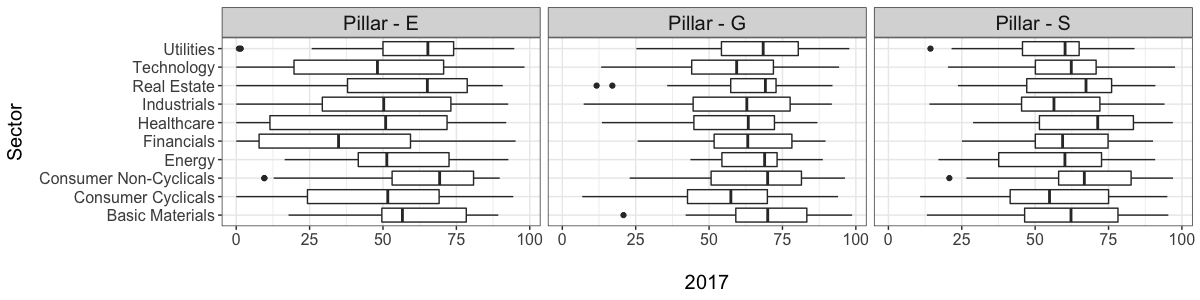}
   \label{fig:data-17}
\end{figure} 

We estimate the companies' annual $95\%$ VaR as the empirical quantile and the annual volatility as the market risk measure using the daily logarithmic returns for the risk measures. Then, we calculate the companies' \textit{vvrisk}. Figure 6 in Appendix \ref{app-ratio} reports the pairwise scatter plots with the empirical Kendall's $\tau$ of the VaR, volatility, \textit{vvrisk} for the S\&P 500. We remark that less negative VaR tends to be associated with smaller volatility levels, and the \textit{vvrisk} is highly negatively/positively dependent on the volatility/VaR. Moreover, we can observe the variability in the VaR of the companies in each of the ten sectors in the S\&P 500 and EuroStoxx 600 in Figure 7 in Appendix \ref{app-ratio}, showing that Utilities and Real Estate have smaller variations than others. We present the results using VaR as the risk measure in Equation \eqref{eq:risk-func}. However, the results similarly hold when considering the volatility or \textit{vvrisk} in Equation \eqref{eq:risk-func} (available upon request). 

Next, we report the descriptive statistics of the companies' M pillar scores in Section \ref{m-pillar-emp}. Since we have three years ($t \in \{2017,2018,2019\}$) in our data sets, we use 2017 and 2018 as the in-sample data to estimate the pillar weights in Section \ref{in-sample}. Later, in Section \ref{out-of-sample}, we apply our estimated pillar weights to the out-of-sample data, i.e., E, S, G, and M pillar scores of 2019, to calculate the ESGM scores for 2019. Then, we compare their risk performance with the ESG scores for 2019 as an out-of-sample analysis. We run separate analyses for the S\&P 500 and EuroStoxx 600 and discuss our results in Section \ref{disc}.

\subsection{M pillar scores}\label{m-pillar-emp} 
Since the companies in both data sets contain the ESG categories with not yet reported ESG information as shown in Table 1, we account for their information disclosure, assigning an M pillar score as in Equation \eqref{eq:m-pillar} in each of the ten sectors. The number of ESG categories with not yet disclosed ESG information varies from zero to six as reported in Appendix \ref{app-missing-esg}. EuroStoxx 600 companies have less missing information in the ESG categories than the S\&P 500 companies. Given that there are mandatory ESG disclosure regulations in European Union, but not in the USA, such observations might be expected \parencite{EuropeanCom}.  Moreover, some sectors such as Utilities in the EuroStoxx 600, have more ESG disclosures than others in the same geographical region. Thus, missing ESG information changes with sectors and geographical regions.

After computing the companies' M pillar scores across ten sectors and three years in the S\&P 500 and EuroStoxx 600, we observe that the mean M pillar score within each sector and year in both data sets is 50, as constructed. The M pillar score shows variation among ten sectors, and the standard deviation of the M pillar score changes from 18.31 for Consumer Non-Cyclicals to 28.16 for Real Estate in 2017 in the S\&P 500. Likewise, the lowest and highest M pillar standard deviations are 15.75 for Utilities in 2019 and 28.28 for Real Estate in 2017 in the EuroStoxx 600. Consumer Non-Cyclicals and Utilities have the lowest percentage of the companies with undisclosed ESG information regarding an ESG category in Table 1 for the S\&P 500 and EuroStoxx 600, respectively, and their M pillar score has the lowest standard deviation.

\newpage
Table 2 reports the empirical Kendall's $\tau$ values between the ESG scores, E, S, G, and M pillar scores in Consumer Cyclicals in 2017 in the S\&P 500. We see that the E, S, and G pillar scores have positive medium-sized dependence on the ESG scores, while the M pillar score negatively depends on the ESG scores and the other pillars, as expected. When more ESG information is available for a company, the number of ESG categories with undisclosed ESG information decreases. Accordingly, its ESG category scores increase, assuming nothing changes in the other available information. Then, since an ESG pillar score aggregates the underlying ESG category scores, the respective E, S, and G pillar scores increase, increasing its ESG score. However, since the number of zero values used for the computation of the M pillar score decreases, its M pillar score goes down. Our findings are characteristically similar when considering other years, sectors, and the data set EuroStoxx 600.

\begin{table}[H]
\centering
\begin{tabular}{lrrrrr}
  \hline
 & ESG & E & S  & G & M \\ 
  \hline
ESG & 1.00 &  &  &  &  \\ 
E & 0.66 & 1.00  &  &  &  \\ 
 S & 0.76 & 0.54 & 1.00 &  &  \\ 
G & 0.46 & 0.25 & 0.31 & 1.00 &  \\ 
M  & -0.52 & -0.57 & -0.44 & -0.20 & 1.00 \\ 
   \hline
\end{tabular}
\caption{Empirical Kendall's $\tau$ matrix of the ESG scores, E, S, G, and M pillar scores in Consumer Cyclicals in S\&P 500 in 2017.}
\label{table2}
\end{table}  

\subsection{In-sample analysis}\label{in-sample}
After computing the companies' E, S, G, and M pillar scores, now, we focus on estimating the E, S, G, and M pillar score weights using the data in 2017 and 2018 across ten sectors, linking the resulting ESGM scores to their risk measures as formulated in Equation \eqref{eq:opt-problem-sec-esg}. Precisely, we find $(\hat{w}^{a}_{E}, \hat{w}^{a}_{S},\hat{w}^{a}_{G}, \hat{w}^{a}_{M})$ for all sectors $a$ in both data sets and aim to analyze for which sectors there is a risk strengthening effect using the M pillar, i.e., potential disclosure of ESG information.

Table 3 reports the sectors' estimated four pillar score weights, where the M pillar weight is not zero. For the remaining sectors, where the M pillar weight is zero, the re-weighting scheme for the E, S, and G pillar scores usually leads to stronger risk dependence than the original ESG scores as given in Appendix \ref{app-m-zero}. According to Table 3, as postulated in Equation \eqref{eq:opt-problem-sec-esg}, the pillar score weights sum up to one; the E, S, and G pillar score weights are at least 0.100; the M pillar score is non-negative, and the E, S, and G pillar score weights are larger than or equal to the M pillar score weight in both data sets. We also see that ESGM scores are built on the M pillar (non-zero M pillar weight) for Consumer Cyclicals, Energy, Industrials, Technology, and Utilities in the S\&P 500 and Consumer Cyclicals, Energy, Industrials, Financials, Healthcare, and Real Estate in the EuroStoxx 600. Even though we observe the M pillar effect on the risk dependence in Consumer Cyclicals, Energy, Industrials in the S\&P 500 and EuroStoxx 600, the estimated pillar weights differ in both data sets. For instance, the M pillar weight for Consumer Cyclicals is 0.084 in the S\&P 500 and 0.189 in the EuroStoxx 600. Moreover, Technology has the M pillar weight of 0.245 in the S\&P 500 and of zero in the EuroStoxx 600. Therefore, the impact of the potential disclosure of ESG information, i.e., missing ESG information, on the VaR dependence changes by sectors and geographical regions.

\begin{table}[H]
\centering
\setlength{\tabcolsep}{2pt}
\begin{tabular}{lcccc}
\hline
Sector (S\&P)               & E     & S     & G      & M \\
\hline
  Consumer Cycl. & 0.259 & 0.195 & 0.357 & 0.189 \\ 
  Energy & 0.650 & 0.100 & 0.172 & 0.078 \\ 
  Industrials & 0.102 & 0.103 & 0.751 & 0.044 \\ 
  Technology & 0.245 & 0.245 & 0.265 & 0.245 \\ 
  Utilities & 0.323 & 0.323 & 0.177 & 0.177 \\ 
             &  &     &     &  \\
\hline
\end{tabular} 
\quad
\begin{tabular}{lcccc}
\hline
Sector (EuroStoxx)         & E     & S     & G      & M \\
\hline
  Consumer Cycl. & 0.240 & 0.541 & 0.135 & 0.084 \\ 
  Energy & 0.238 & 0.485 & 0.175 & 0.102 \\ 
    Industrials & 0.100 & 0.714 & 0.100 & 0.086 \\ 
  Financials & 0.100 & 0.100 & 0.700 & 0.100 \\ 
 Healthcare & 0.236 & 0.562 & 0.101 & 0.101 \\ 
  Real Estate & 0.700 & 0.100 & 0.100 & 0.100 \\ 
\hline
\end{tabular}
 \caption{New E, S, G, M pillar score weights, resulting in the ESGM scores in Tables 4 and 5 across sectors for which we have a non-zero M pillar score weight in S\&P 500 (left) and EuroStoxx 600 (right).}
\label{table3}
\end{table} 

Remarkably, the dependence of ESG scores and risk also depends on sectors and geographical regions as shown in Table 4. While the dependence on Industrials is significant at the 10\% level with the value of 0.118 in 2017 in the S\&P 500, it is 0.060 without being significant at the same level in the EuroStoxx 600 (\cite{Hollander2014}). Nevertheless, in Table 4, the ESGM scores provide stronger VaR dependence for the sectors with a non-zero M pillar weight in both data sets. 

\begin{table}[H]
\centering
    \begin{tabular}{lrlrl}
    \hline
     & \multicolumn{4}{c}{Panel A: ESG and VaR} \\
      \hline
Sector (S\&P) &2017 &&2018   \\
\hline
Consumer Cycl. &0.161&**& 0.144&**  \\
Energy           &   0.196 &*& 0.203&*  \\
Industrials      &   0.118 &*&-0.071&  \\
Technology    &      0.154 &**& 0.028&  \\
Utilities       &   -0.079 &&-0.030&  \\
\hline
Sector (EuroStoxx) &2017 &&2018   \\
\hline
Consumer Cycl.  &     0.144 &**& 0.157&**  \\
Energy             &      0.471 &***& 0.515&***  \\
Industrials       &       0.060 && 0.029&  \\
Financials      &        -0.132 &&-0.088&  \\
Healthcare   &            0.205  &**&0.432&***  \\
Real Estate   &          -0.040&& -0.151&  \\
\hline
  \end{tabular}
      \begin{tabular}{rlrl}
      \hline
      \multicolumn{4}{c}{Panel B: ESGM and VaR} \\
      \hline
2017 &&2018   \\
\hline
  0.216 &***& 0.215&***  \\
       0.261  &**&0.355& *** \\
     0.137 &*&0.019&  \\
    0.189 &***& 0.094&  \\
       0.015  &&0.177&*  \\
\hline
2017 &&2018   \\
\hline
     0.187&***&  0.194& *** \\
            0.559  &***&0.574&***  \\
                        0.094 && 0.070&  \\
         -0.059 && 0.001&  \\
            0.258 &**& 0.466&***  \\
           -0.015 &&-0.114&  \\
           \hline
  \end{tabular}
    \caption{Kendall's $\tau$ between ESG, ESGM scores, and 95\% VaR in 2017, 2018 across sectors (in-sample) for which we have a non-zero M pillar score weight in S\& P500 (top) and EuroStoxx 600 (bottom). The hypothesis testing is $H_0:\tau=0$ versus $H_A:\tau>0$. *, **, and *** denote statistical significance at 10\%, 5\%, and 1\% levels, respectively.}
    \label{table4}
\end{table}  

A company is assigned to a rating class (i.e., A, B, C, or D) based on its ESG score using thresholds or quartiles (e.g., \cite{Refinitiv2021}). Thus, we group the companies with the highest to lowest ESG and ESGM scores in the first/second/third/fourth quartile as ESG and ESGM rating class A/B/C/D in each of the sectors where we observe the M pillar effect, respectively. Class D contains the companies with the lowest scores and might be excluded from ESG portfolios. For both data sets, we witness that low ESGM scores (class D) are associated with higher or equal median risks than low ESG scores (class D), except for Industrials in 2017 in the EuroStoxx 600, as demonstrated in Figure 4. Nonetheless, the median risk of ESG and ESGM scores in the EuroStoxx 600 is closer than that of those in the S\&P500.

Additionally, since the VaR variation is low in some sectors, such as Utilities in the S\&P 500, dividing the companies into classes with different VaR characteristics is hard. Thus, we can argue that the companies which have not yet released ESG information as much as their peers do, thereby having lower ESG scores than them in some sectors, might result in risk underestimation in the ESG portfolios using negative screening. Instead, the ESGM scores quantify better the companies that can be excluded, e.g., ESGM class D, and provide stronger risk performances for such portfolios than the ESG scores as seen for Consumer Cyclicals in the S\&P 500 in 2017 and Healthcare in the EuroStoxx 600 in 2018. 

\begin{figure}[H]
\centering
     \caption{Empirical 95\% VaR of rating class D for ESG and ESGM in S\&P 500 (left) and EuroStoxx 600 (right) in 2017, 2018.}
\includegraphics[width=.49\textwidth]{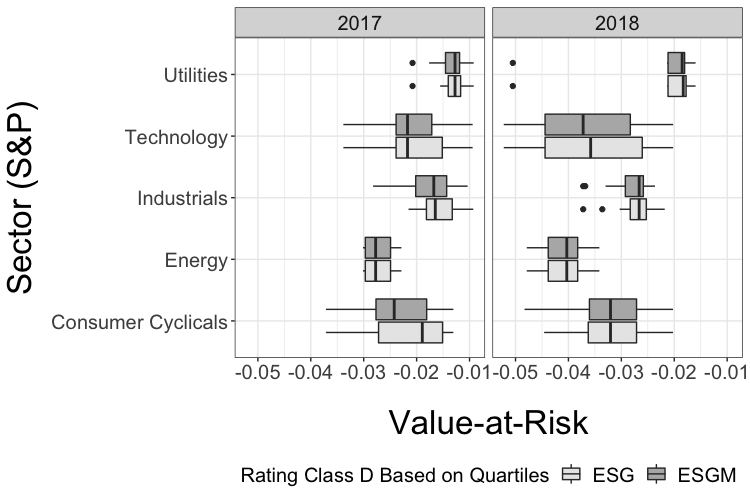} \hfill
\includegraphics[width=.49\textwidth]{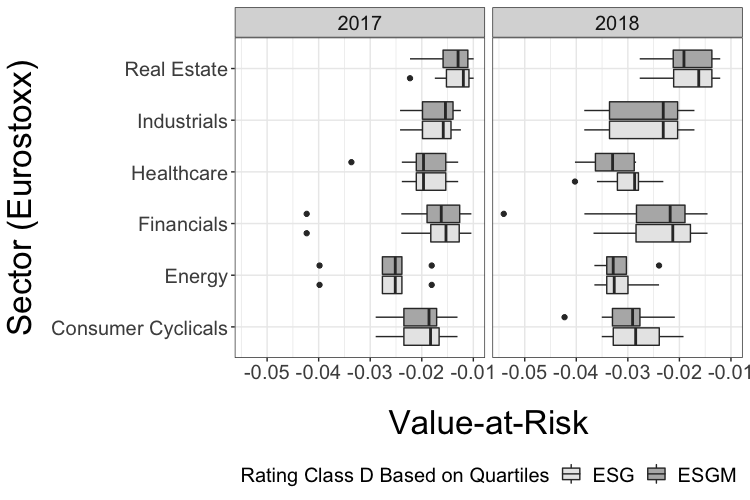}
   \label{fig:esg-esgm-in}
\end{figure}

Comparing the ESG and ESGM rating classes in Consumer Cyclicals in the S\&P 500 in 2017 presents that ESGM scores move three/one companies from the ESG class D to the ESGM class C/B in 2017. Likewise, one company in the ESGM class B and three companies in the ESGM class C belong to the ESG class D in Healthcare in the EuroStoxx 600 in 2018. Such results reveal that the companies with low ESG scores might not necessarily provide the worst risk performances. Rather, their ESG scores could be low due to not yet disclosed ESG information the data provider does not explicitly point out. Still, these companies might publish more ESG information in the future, increasing their ESG scores, as modeled by their ESGM scores. Therefore, the ESGM scores can work not only to include missing information but also to allocate the companies in more appropriate risk classes. Similar results hold for the remaining sectors in both data sets with a non-zero M pillar weight (available upon request).

\subsection{Out-of-sample analysis}\label{out-of-sample}
ESGM scores provide stronger risk dependence and identify more reliably the companies for exclusion strategies in ESG portfolios than the ESG scores as discussed in Section \ref{in-sample}. We use the estimated E, S, G, and M pillar score weights for another year to calculate ESGM scores, where the ESG data is still non-definitive, and companies can publish their ESG information in time. Accordingly, we perform the out-of-sample analyses using the data sets for 2019. 

First, we calculate the companies' ESGM scores for 2019 in each sector and data set as follows:
\begin{align*}
x^{a}_{ESGM, p, 2019} = x^{a}_{E, p, 2019} \cdot \hat{w}^{a}_{E} + x^{a}_{S, p, 2019} \cdot \hat{w}^{a}_{S} + x^{a}_{G, p, 2019} \cdot \hat{w}^{a}_{G} + x^{a}_{M, p, 2019} \cdot \hat{w}^{a}_{M}, \forall{a,p},\\
\end{align*}
where $\hat{w}^{a}_{E}$, $\hat{w}^{a}_{S}$, $\hat{w}^{a}_{G}$, and $\hat{w}^{a}_{M}$ are the estimated weights using the training data in 2017 and 2018 in Section \ref{in-sample}.

Second, we analyze the dependence between the ESG scores, ESGM scores and VaR in 2019 in Table 5, again focusing on the sectors analyzed in our in-sample-analysis for both data sets in Section \ref{in-sample}. As can be seen, the out-of-sample analysis also confirms the higher risk dependence for the ESGM scores than the ESG scores in all cases but Energy in the EuroStoxx 600. However, the significance of the risk dependence is better using the ESGM scores than the ESG scores in Energy in the S\&P 500. 

\begin{table}[H]
\centering
    \begin{tabular}{lrlrl}
    \hline
     & \multicolumn{3}{c}{Panel A: ESG and VaR} \\
      \hline
Sector (S\&P500) &2019 &  \\
\hline
Consumer Cycl. &0.036&\\
Energy           &0.229&*\\
Industrials     &-0.103&\\
Technology  &0.051&\\
Utilities      &-0.099&\\
\hline
Sector (EuroStoxx) &2019 &  \\
\hline
Consumer Cycl. &  -0.019&  \\
Energy               &   0.529&***  \\
Industrials      &      0.030&  \\
Financials       &      -0.123&  \\
Healthcare     &         0.375&***  \\
Real Estate     &       -0.182&  \\
\hline
  \end{tabular}
      \begin{tabular}{rlrl}
      \hline
      \multicolumn{2}{c}{Panel B: ESGM and VaR} \\
      \hline
2019 &   \\
\hline
0.096&   \\
0.326& **  \\
0.039&   \\
0.068&   \\
-0.010&   \\
\hline
2019 &   \\
\hline
 0.015&  \\
  0.529& ***\\
   0.076&  \\
-0.031&  \\
 0.386&***  \\
  -0.120&  \\
\hline
  \end{tabular}
      \caption{Kendall's $\tau$ between ESG, ESGM scores, and 95\% VaR in 2019 across sectors (out-of-sample) for which we have a non-zero M pillar score weight in S\&P 500 (top) and EuroStoxx 600 (bottom).} 
          \label{table5}
\end{table}

Finally, Figure 5 suggests that the ESGM class D presents higher median risks than the ESG class D in all but Utilities in the S\&P 500. Similarly, the median risk ESGM scores provided for class D is higher than that of ESG scores provided for Consumer Cyclicals, Financials, and Real Estate in the EuroStoxx 600. However, the median risk of the ESG and ESGM scores for class D seems to be closer in the EuroStoxx 600 than the S\&P 500. Hence, the ESG portfolios using negative screening could benefit from the ESGM scores in terms of the risk performance, supporting our findings using the in-sample data in Section \ref{in-sample}, even though there are some differences across sectors and regions.

\begin{figure}[H]
\centering
     \caption{Empirical 95\% VaR of rating class D for ESG and ESGM in S\&P 500 (left) and EuroStoxx 600 (right) in 2019.}
\includegraphics[width=.49\textwidth]{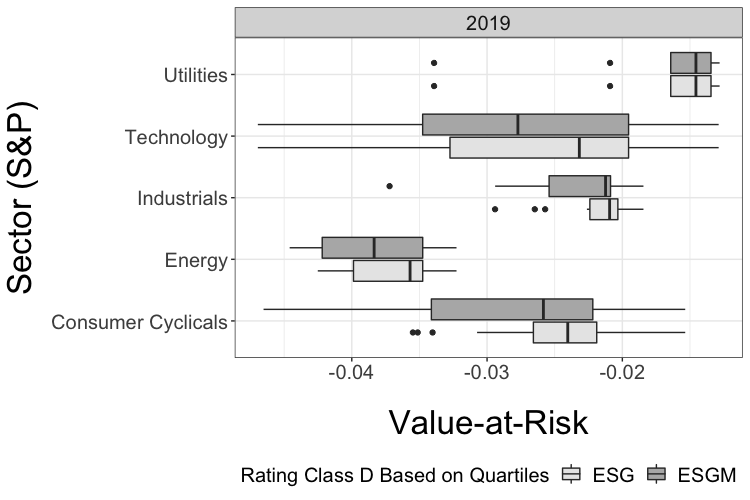} \hfill
\includegraphics[width=.49\textwidth]{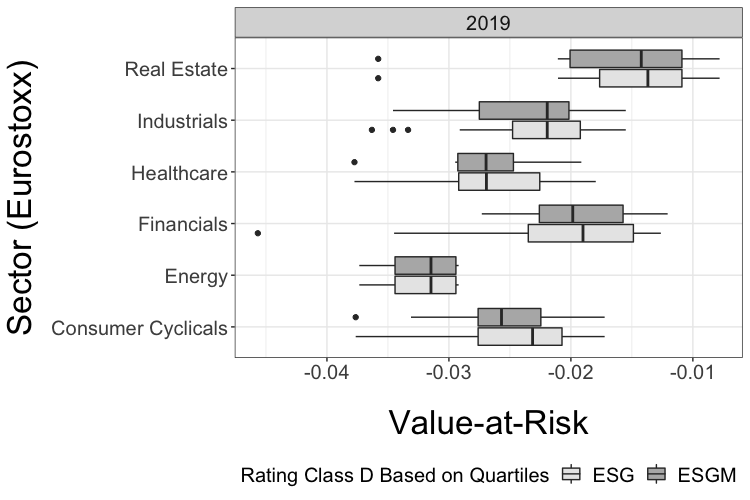}
   \label{fig:esg-esgm-out}
\end{figure}

\subsection{Discussion}\label{disc}
Our findings support the proposed methodology and provide evidence of the need to consider disclosing new ESG information that can be linked to the companies' riskiness and capital misallocation. Our research is in alignment with the recent discussions. The Sustainable Finance Roadmap, released in February 2022 by European Securities and Markets Authority (ESMA), lists the main challenges which need action in analyzing the ESG-related risks. They report that \say{Overall, data gaps, low quality and a lack of transparency may lead to misrepresentation and to a misallocation/mispricing of investments} and call for actions to assess the issue of the data quality affecting the ESG data users (Page 19 of \cite{ESMA2022}).

Our empirical findings provide that the non-financial disclosures are higher in the EuroStoxx 600 companies than in the S\&P 500 companies. In the European Union, sustainable finance legislations have been fine-tuned. For instance, the European Union's (EU) Directive 2014/95/EU sets out that relevant, large, and public interest EU companies must disclose ESG information annually.\footnote{\url{https://eur-lex.europa.eu/legal-content/EN/TXT/?uri=CELEX\%3A32014L0095}} On the other hand, there have not yet existed any mandatory sustainability-related disclosures in the United States \parencite{EuropeanCom}. However, the Federal Reserve joined the Network for Greening the Financial System, a global network of central banks, and might take action about the applicable commitments towards a sustainable economy.\footnote{\url{https://www.federalreserve.gov/newsevents/pressreleases/bcreg20201215a.htm}} The mandatory context might have a positive impact on the credibility of the non-financial disclosures for the companies \parencite{Mazzotta2020}.

Our estimated M pillar weight, which shows the impact of not yet released ESG information on the risk dependence, is zero for some sectors despite having many companies with not yet disclosed ESG information for at least one of the ten ESG categories. It implies that the missing information's potential impact on the risk differs for sectors and geographical regions. For researchers, this suggests investigating the determinants of the occurrence and distribution of the companies' missing ESG information among ESG categories across regions and sectors.  Such an analysis could also provide insights into which and how ESG information should be disclosed to measure the companies' ESG performance and responsible investing accurately. Recently, ESMA also stated regarding the ESG data that \say{These data needs are currently not fulfilled by the data disclosed by companies. The data gaps can neither be fully bridged by third-party ESG data or by rating providers whose methodologies, limitations and assumptions need to become more transparent.} (Page 19 of \cite{ESMA2022}). \cite{Lagasio2019} already identified board independence, the board size, and women's directorship as the empowering factors of non-financial disclosures. However, as they pointed out, there is a gap in the studies investigating the determinants of the non-financial information released by companies across sectors.

Furthermore, for researchers, in addition to showing companies' potential to disclose missing ESG information, the M pillar can be used as a proxy of the current ESG disclosure quality adapted for sectoral peculiarities like other pillars. Future studies can perform a regression analysis, where the dependent variable is a financial performance measure, and independent variables are E, S, G, and M pillar scores and company specifics like their market capitalization. This would allow analyzing the impact of the current disclosure quality on financial measures. However, we remark that such an analysis does not provide scores comparable to ESG scores; thus, they might be difficult to be used by investors in the same manner. Alternative approaches for a regression analysis can encode the current disclosure quality as a binary variable based on the industry median by counting the number of ESG categories disclosed as proposed in \cite{Santamaria2021}.

Our results show the importance of knowing and understanding what is behind the ESG scores for investors. The investors can exclude the companies based on their low ESG scores in a given sector from their portfolios; however, \cite{Sahin2022} reported that the companies still disclose their ESG information in time, increasing their ESG scores.  Additionally, we discuss that using the proposed ESGM scores provides better risk performances than the ESG scores for such cases. 

For companies and managers, our analyses imply that they could be excluded from investment portfolios, not necessarily because of their actual ESG performance but possibly due to their lack of and low speed of the ESG disclosure mechanism. Therefore, a focus should be on providing complete disclosure material, allowing investors to manage the ESG-related risks better and make sustainable investments for the world.

Finally,  we like to remark that while our results are based on the missing data  of the single data provider, Refinitiv, another provider, Sustainalytics, also encodes the missing raw scores based on company disclosures as zero.\footnote{\url{https://www.sustainalytics.com/docs/default-source/meis/kone_oyj_riskratingsreport_18032021.pdf?sfvrsn=577d22c8_0 }} Yet, the applications of the M pillar and ESGM scores using different providers are left for future research.

\section{Conclusion}\label{conc}
We propose a new pillar score, the so-called Missing (M-) pillar score, to explicitly consider the companies’ potential of disclosing missing ESG information. By doing so, we introduce a new Environmental, Social, Governance, and Missing (ESGM) score. In addition, our study formulates an optimization scheme to link the companies’ ESGM scores and riskiness. Such a scheme encourages companies to disclose more ESG information and evaluate the impact of such additional information on financial risk.  Furthermore, it  can be used by investors to build customized scores.

We evaluate the risk performance of the proposed ESGM scores and ESG scores using in-sample and out-of-sample data. The in-sample data analysis allows us to estimate the ESGM scores' pillar weights, which are used to compute the ESGM scores. The out-of-sample data analysis tests the power of the ESGM scores regarding their risk dependence. Using the S\&P 500 and EuroStoxx 600 companies' non-definitive ESG data provided by Refinitiv, we show that the ESGM scores provide stronger risk dependence than the ESG scores in some sectors using both the in-sample and out-of-sample data. We argue that a potential disclosure of missing ESG information impacts the risk in these sectors. This approach potentially supports the investment decisions as an ESG exclusion strategy depends on the ESG rating class, which is affected by missing information. Furthermore, incorporating the potential of possible disclosure allows to include companies in the portfolio that would otherwise be excluded as too much data is missing at the time of the investment decision. Nonetheless, the dependence of risk and ESG/ESGM scores and the impact and amount of the missing ESG information change with sectors and geographical regions.

Our study is limited by a small number of companies within each sector. Future studies should consider applying our M pillar score and ESGM scores formulations to data from a larger number of companies. The second limitation of our research is that it does not account for common risk factors. Moreover, necessary modifications for the M pillar score can be considered in the future to evaluate the ESGM scores also from other ESG data providers. Lastly, exploring optimization schemes with different objective functions and constraints can be considered in the future.
\newpage
\printbibliography

@misc{EuropeanCom,
author = {{International Platform on Sustainable Finance (IPSF)}},
title = {{International Platform on Sustainable Finance}},
howpublished ={\url{https://ec.europa.eu/info/sites/default/files/business_economy_euro/banking_and_finance/documents/211104-ipsf-esg-disclosure-report_en.pdf}},
note = {visited on 2022-04-20},
year = {2021}
}

@article{Abhayawansa2021,
  title={{Sustainable investing: The black box of environmental, social, and governance (ESG) ratings}},
  author={Abhayawansa, Subhash and Tyagi, Shailesh},
  journal={The Journal of Wealth Management},
  volume={24},
  number={1},
  pages={49--54},
  year={2021},
  publisher={Institutional Investor Journals Umbrella}
}

@article{Ahmed2021,
  title={{Modeling demand for ESG}},
  author={Ahmed, Muhammad Farid and Gao, Yang and Satchell, Stephen},
  journal={The European Journal of Finance},
  pages={1--15},
  year={2021},
  publisher={Taylor \& Francis}
}

@article{Alessandrini2020,
  title={{ESG investing: From sin stocks to smart beta}},
  author={Alessandrini, Fabio and Jondeau, Eric},
  journal={The Journal of Portfolio Management},
  volume={46},
  number={3},
  pages={75--94},
  year={2020},
  publisher={Institutional Investor Journals Umbrella}
}

@article{Alessandrini2021,
  title={{ESG Screening in the Fixed-Income Universe}},
  author={Alessandrini, Fabio and Baptista Balula, David and Jondeau, Eric},
  journal={Available at SSRN 3966312},
  doi={10.2139/ssrn.3966312},
  year={2021}
}

@article{Amel2018,
  title={{Why and how investors use ESG information: Evidence from a global survey}},
  author={Amel-Zadeh, Amir and Serafeim, George},
  journal={Financial Analysts Journal},
  volume={74},
  number={3},
  pages={87--103},
  year={2018},
  doi={10.2469/faj.v74.n3.2},
  publisher={Taylor \& Francis}
}

@article{Bax2021,
title={{ESG, Risk, and (Tail) Dependence}},
author={Bax, Karoline and Sahin, {\"O}zge and Czado, Claudia and Paterlini, Sandra},
year={2021},
journal={Available at SSRN 3846739},
doi={10.2139/ssrn.3846739}
}

@article{Behl2021,
  title={{Exploring the relationship of ESG score and firm value using cross-lagged panel analyses: Case of the Indian energy sector}},
  author={Behl, Abhishek and Kumari, PS and Makhija, Harnesh and Sharma, Dipasha},
  journal={Annals of Operations Research},
  pages={1--26},
  year={2021},
  publisher={Springer}
}

@article{Berg2019,
title={{Aggregate Confusion: The Divergence of ESG Ratings}},
author={Berg, Florian and K{\"o}lbel, Julian F and Rigobon, Roberto},
journal={Available at SSRN 3438533},
doi = {10.2139/ssrn.3438533},
year={2020}
}

@article{Berg2021,
author = {Berg, Florian and Fabisik, Kornelia and Sautner, Zacharias},
title = {{Is History Repeating Itself? The (Un)Predictable Past of ESG Ratings}},
doi = {110.2139/ssrn.3722087},
journal={Available at SSRN 3722087},
year = {2021}
}

@article{Billio2021,
author = {Billio, Monica and Costola, Michele and Hristova, Iva and Latino, Carmelo and Pelizzon, Loriana},
title = {{Inside the ESG ratings: (Dis)agreement and performance}},
journal = {Corporate Social Responsibility and Environmental Management},
volume = {28},
number = {5},
pages = {1426-1445},
year = {2021}
}

@article{Cornett2016,
title = {{Greed or good deeds: An examination of the relation between corporate social responsibility and the financial performance of U.S. commercial banks around the financial crisis}},
journal = {Journal of Banking \& Finance},
volume = {70},
pages = {137-159},
year = {2016},
issn = {0378-4266},
author = {Marcia Millon Cornett and Otgontsetseg Erhemjamts and Hassan Tehranian}
}

@article{Diemont2016,
  title={{The downside of being responsible: Corporate social responsibility and tail risk}},
  author={Diemont, Dolf and Moore, Kyle and Soppe, Aloy},
  journal={Journal of Business Ethics},
  volume={137},
  number={2},
  pages={213--229},
  year={2016},
  doi={10.1007/s10551-015-2549-9},
  publisher={Springer}
}

@article{Fatemi2018,
  title={{ESG performance and firm value: The moderating role of disclosure}},
  author={Fatemi, Ali and Glaum, Martin and Kaiser, Stefanie},
  journal={Global Finance Journal},
  volume={38},
  pages={45--64},
  year={2018},
  publisher={Elsevier}
}

@article{Friede2015,
  title={{ESG and financial performance: aggregated evidence from more than 2000 empirical studies}},
  author={Friede, Gunnar and Busch, Timo and Bassen, Alexander},
  journal={Journal of Sustainable Finance \& Investment},
  volume={5},
  number={4},
  pages={210--233},
  year={2015},
  publisher={Taylor \& Francis}
}

@article{Giese2021,
  title={{Deconstructing ESG Ratings Performance: Risk and Return for E, S, and G by Time Horizon, Sector, and Weighting}},
  author={Giese, Guido and Nagy, Zolt{\'a}n and Lee, Linda-Eling},
  journal={The Journal of Portfolio Management},
  volume={47},
  number={3},
  pages={94--111},
  year={2021},
  doi = {10.3905/jpm.2020.1.198},
  issn = {0095-4918},
  publisher={Institutional Investor Journals Umbrella}
}

@article{Ghoul2017,
title = {{Does corporate social responsibility affect mutual fund performance and flows?}},
journal = {Journal of Banking \& Finance},
volume = {77},
pages = {53-63},
year = {2017},
issn = {0378-4266},
doi = {https://doi.org/10.1016/j.jbankfin.2016.10.009},
author = {Sadok {El Ghoul} and Aymen Karoui}
}

@article{Gyonyorova2021,
  title={{ESG ratings: relevant information or misleading clue? Evidence from the S\&P Global 1200}},
  author={Gy{\"o}ny{\"o}rov{\'a}, Lucie and Stacho{\v{n}}, Martin and Sta{\v{s}}ek, Daniel},
  journal={Journal of Sustainable Finance \& Investment},
  pages={1--35},
  year={2021},
  doi={10.1080/20430795.2021.1922062},
  publisher={Taylor \& Francis}
}

@article{Henke2016,
title = {{The effect of social screening on bond mutual fund performance}},
journal = {Journal of Banking \& Finance},
volume = {67},
pages = {69-84},
year = {2016},
issn = {0378-4266},
author = {Hans-Martin Henke}
}

@article{Hou2019,
  title={{Does CSR matter? Influence of corporate social responsibility on corporate performance in the creative industry}},
  author={Hou, Chen-En and Lu, Wen-Min and Hung, Shiu-Wan},
  journal={Annals of Operations Research},
  volume={278},
  number={1},
  pages={255--279},
  year={2019},
  publisher={Springer}
}

@article{Kalaitzoglou2021,
  title={{Corporate social responsibility: How much is enough? A higher dimension perspective of the relationship between financial and social performance}},
  author={Kalaitzoglou, Iordanis and Pan, Hui and Niklewski, Jacek},
  journal={Annals of Operations Research},
  volume={306},
  number={1},
  pages={209--245},
  year={2021},
  doi={10.1007/s10479-020-03834-y},
  publisher={Springer}
}

@article{Kudratova2020,
  title={{Corporate sustainability and stakeholder value trade-offs in project selection through optimization modeling: Application of investment banking}},
  author={Kudratova, Shamsiya and Huang, Xiaoxia and Kudratov, Khikmatullo and Qudratov, Shohrukh},
  journal={Corporate Social Responsibility and Environmental Management},
  volume={27},
  number={2},
  pages={815--824},
  year={2020},
  doi={10.1002/csr.1846},
  publisher={Wiley Online Library}
}

@article{Kumar2016,
author = {N. C. Ashwin Kumar and Camille Smith and Le{\"\i}la Badis and Nan Wang and Paz Ambrosy and Rodrigo Tavares},
title = {{ESG factors and risk-adjusted performance: a new quantitative model}},
journal = {Journal of Sustainable Finance \& Investment},
volume = {6},
number = {4},
pages = {292-300},
year  = {2016},
doi = {10.1080/20430795.2016.1234909},
publisher={Taylor \& Francis}
}

@article{Lagasio2019,
  title={{Corporate governance and environmental social governance disclosure: A meta-analytical review}},
  author={Lagasio, Valentina and Cucari, Nicola},
  journal={Corporate Social Responsibility and Environmental Management},
  volume={26},
  number={4},
  pages={701--711},
  year={2019},
  doi={10.1002/csr.1716},
  publisher={Wiley Online Library}
}

@article{Larson2019, 
title={Derivative-free optimization methods}, 
volume={28}, 
DOI={10.1017/S0962492919000060}, 
journal={Acta Numerica},
publisher={Cambridge University Press}, 
author={Larson, Jeffrey and Menickelly, Matt and Wild, Stefan M.}, 
year={2019}, 
pages={287–404}
}

@article{Maiti2021,
author = {Moinak Maiti},
title = {Is ESG the succeeding risk factor?},
journal = {Journal of Sustainable Finance \& Investment},
volume = {11},
number = {3},
pages = {199-213},
year  = {2021},
publisher = {Taylor & Francis},
doi = {10.1080/20430795.2020.1723380}
}

@article{Mazzotta2020,
  title={{Are mandatory non-financial disclosures credible? Evidence from Italian listed companies}},
  author={Mazzotta, Romilda and Bronzetti, Giovanni and Veltri, Stefania},
  journal={Corporate Social Responsibility and Environmental Management},
  volume={27},
  number={4},
  pages={1900--1913},
  year={2020},
  doi={10.1002/csr.1935},
  publisher={Wiley Online Library}
}

@article{Pedersen2021,
  title={{Responsible investing: The ESG-efficient frontier}},
  author={Pedersen, Lasse Heje and Fitzgibbons, Shaun and Pomorski, Lukasz},
  journal={Journal of Financial Economics},
  volume={142},
  number={2},
  pages={572--597},
  year={2021},
  doi={10.1016/j.jfineco.2020.11.001},
  publisher={Elsevier}
}

@article{Powell2015,
author = {Powell, M. J. D.},
doi = {10.1007/s12532-015-0084-4},
journal = {Mathematical Programming Computation},
pages = {237--267},
volume = {7},
title = {{On fast trust region methods for quadratic models with linear constraints}},
year = {2015}
}

@article{Sahin2022,
title={{The pitfalls of (non-definitive) Environmental, Social, and Governance scoring methodology}},
author={Sahin, {\"O}zge and Bax, Karoline and Paterlini, Sandra and Czado, Claudia},
doi = {10.2139/ssrn.4020354},
journal={Available at SSRN 4020354},
year = {2022}
}

@article{Santamaria2021,
  title={{Non-financial strategy disclosure and environmental, social and governance score: Insight from a configurational approach}},
  author={Santamaria, Riccardo and Paolone, Francesco and Cucari, Nicola and Dezi, Luca},
  journal={Business Strategy and the Environment},
  volume={30},
  number={4},
  pages={1993--2007},
  year={2021},
  doi={10.1002/bse.2728},
  publisher={Wiley Online Library}
}

@article{Shafer2020,
  title={Environmental, social, and governance practices and perceived tail risk},
  author={Shafer, Michael and Szado, Edward},
  journal={Accounting \& Finance},
  volume={60},
  number={4},
  pages={4195--4224},
  year={2020},
  publisher={Wiley Online Library}
}

@article{Verheyden2016,
author = {Verheyden, Tim and Eccles, Robert G. and Feiner, Andreas},
title = {{ESG for All? The Impact of ESG Screening on Return, Risk, and Diversification}},
journal = {Journal of Applied Corporate Finance},
volume = {28},
number = {2},
pages = {47-55},
doi = {https://doi.org/10.1111/jacf.12174},
year = {2016}
}

@article{Zhang2021,
  title={{Implied Tail Risk and ESG Ratings}},
  author={Zhang, Jingyan and De Spiegeleer, Jan and Schoutens, Wim},
  journal={Mathematics},
  volume={9},
  number={14},
  pages={1611},
  year={2021},
  doi={10.3390/math9141611}
}

@book{Hollander2014,
title={Nonparametric statistical methods},
author={Hollander, Myles and Wolfe, Douglas A and Chicken, Eric},
volume={751},
year={2013},
booktitle = {{Wiley Series in Probability and Statistics}},
doi = {10.1002/9781119196037},
publisher={John Wiley \& Sons}
}

@misc{Bloomberg,
  title = {{ESG assets may hit 53 trillion dollars by 2025, a third of global AUM}},
  author = {Bloomberg},
  howpublished = {\url{https://www.bloomberg.com/professional/blog/esg-assets-may-hit-53-trillion-by-2025-a-third-of-global-aum/}},
  note = {visited on 2021-11-22},
  year = {2021}
}

@misc{Pri,
  title = {{A PRACTICAL GUIDE TO ESG INTEGRATION FOR EQUITY INVESTING}},
  author = {PRI},
  howpublished = {\url{https://www.unpri.org/download?ac=10}},
  note = {visited on 2021-11-22},
  year = {2021}
}

@misc{EBA2021,
title = {{Management and supervision of ESG risks for credit institutions and investment firms}},
author = {EBA},
howpublished = {\url{https://www.eba.europa.eu/eba-publishes-its-report-management-and-supervision-esg-risks-credit-institutions-and-investment}},
note = {visited on 2021-04-26},
year = {2021}
}

@misc{ESMA2022,
title = {{Sustainable Finance Roadmap 2022-2024}},
author = {ESMA},
howpublished = {\url{https://www.esma.europa.eu/sites/default/files/library/esma30-379-1051_sustainable_finance_roadmap.pdf}},
note = {visited on 2022-02-25},
year = {2022}
}

@misc{Refinitiv2021,
author = {Refinitiv},
title = {{Environmental, Social and Governance (ESG) scores from Refinitiv}},
howpublished ={\url{https://www.refinitiv.com/content/dam/marketing/en\_us/documents/methodology/refinitiv-esg-scores-methodology.pdf}},
note = {visited on 2021-10-26},
year = {2021}
}

@misc{Refinitiv2021b,
author = {Refinitiv},
title = {{The Refinitiv Business Classification}},
howpublished = {\url{https://www.refinitiv.com/content/dam/marketing/en\_us/documents/quick-reference-guides/trbc-business-classification-quick-guide.pdf}},
note = {visited on 2021-10-26},
year = {2021}
}

@misc{Refinitiv2021c,
author = {Refinitiv},
title = {{Create Custom ESG Scoring}},
note = {visited on 2022-03-28},
howpublished = {\url{https://www.refinitiv.com/content/dam/marketing/en_us/documents/fact-sheets/build-custom-esg-scores-using-refinitiv-esg-data-in-eikon.pdf}},
year = {2022}
}
\newpage
\appendix

\section{Indices, data, notation}\label{app-notation}
\begin{table}[H]
\centering
\scalebox{0.5}{
\begin{tabular}{ l l}
 Index$\&$Data & Notation \\ 
 \hline
 Sector (Business class)& $a= 1, \ldots, 10$\\ 
 Total number of companies &$n$ \\
 Company $z$ &$z =1, \ldots,n$ \\
 Total number of companies in sector $a$ &$n_a$ \\
  Company $p$ in sector $a$&$p =1, \ldots,n_a$ \\
 Year &  $t =2017,\ldots,2019$\\ \hline
   Value-at-Risk of company $z$ in year $t$ & $x_{VaR, z,t}$ \\
  Value-at-Risk values of companies in year $t$ & $\bm{x}_{VaR, t} = (x_{VaR, 1,t}, \ldots, x_{VaR, n,t})^\top$\\
    Value-at-Risk of company $p$ in sector $a$ and year $t$ & $x^{a}_{VaR, p,t}$ \\
 Value-at-Risk values of companies in sector $a$ and year $t$ & $\bm{x}^{a}_{VaR, t} = (x^{a}_{VaR, 1,t}, \ldots, x^{a}_{VaR, n_a,t})^\top$\\
   Volatility of company $z$ in year $t$ & $x_{vol, z,t}$ \\
  Volatility values of companies in year $t$ & $\bm{x}_{vol, t} = (x_{vol, 1,t}, \ldots, x_{vol, n,t})^\top$\\
    Volatility of company $p$ in sector $a$ and year $t$ & $x^{a}_{vol, p,t}$ \\
 Volatility values of companies in sector $a$ and year $t$ & $\bm{x}^{a}_{vol, t} = (x^{a}_{vol, 1,t}, \ldots, x^{a}_{vol, n_a,t})^\top$\\
    \textit{vvrisk} of company $z$ in year $t$ & $x_{vv, z,t}$ \\
  \textit{vvrisk} values of all companies in year $t$ & $\bm{x}_{vv, t} = (x_{vv, 1,t}, \ldots, x_{vv, n,t})^\top$\\
    \textit{vvrisk} of company $p$ in sector $a$ and year $t$ & $x^{a}_{vv, p,t}$ \\
 \textit{vvrisk} values of companies in sector $a$ and year $t$ & $\bm{x}^{a}_{vv, t} = (x^{a}_{vv, 1,t}, \ldots, x^{a}_{vv, n_a,t})^\top$\\
 ESG score of company $z$ in year $t$ & $x_{ESG, z,t}$ \\
 ESG scores of companies in year $t$ & $\bm{x}_{ESG, t} = (x_{ESG, 1,t}, \ldots, x_{ESG, n,t})^\top$\\
  ESG score of company $p$ in sector $a$ and year $t$ & $x^{a}_{ESG, p,t}$ \\
 ESG scores of companies in sector $a$ and year $t$ & $\bm{x}^{a}_{ESG, t} = (x^{a}_{ESG, 1,t}, \ldots, x^{a}_{ESG, n_a,t})^\top$\\
  ESGM score of company $z$ in year $t$ & $x_{ESGM, z,t}$ \\
 ESGM scores of companies in year $t$ & $\bm{x}_{ESGM, t} = (x_{ESGM, 1,t}, \ldots, x_{ESGM, n,t})^\top$\\
  ESGM score of company $p$ in sector $a$ and year $t$ & $x^{a}_{ESGM, p,t}$ \\
 ESGM scores of companies in sector $a$ and year $t$ & $\bm{x}^{a}_{ESGM, t} = (x^{a}_{ESGM, 1,t}, \ldots, x^{a}_{ESGM, n_a,t})^\top$\\
   E pillar score of company $p$ in sector $a$ and year $t$ & $x^{a}_{E, p,t}$ \\
 E pillar scores of companies in sector $a$ and year $t$ & $\bm{x}^{a}_{E, t} = (x^{a}_{E, 1,t}, \ldots, x^{a}_{E, n_a,t})^\top$\\
   S pillar score of company $p$ in sector $a$ and year $t$ & $x^{a}_{S, p,t}$ \\
 S pillar scores of companies in sector $a$ and year $t$ & $\bm{x}^{a}_{S, t} = (x^{a}_{S, 1,t}, \ldots, x^{a}_{S, n_a,t})^\top$\\
   G pillar score of company $p$ in sector $a$ and year $t$ & $x^{a}_{G, p,t}$ \\
 G pillar scores of companies in sector $a$ and year $t$ & $\bm{x}^{a}_{G, t} = (x^{a}_{G, 1,t}, \ldots, x^{a}_{G, n_a, t})^\top$\\
  M pillar score of company $p$ in sector $a$ and year $t$ & $x^{a}_{M, p,t}$ \\
 M pillar scores of companies in sector $a$ and year $t$ & $\bm{x}^{a}_{M, t} = (x^{a}_{M, 1,t}, \ldots, x^{a}_{M,n_a,t})^\top$\\
   Resource use score of company $p$ in sector $a$ and year $t$ & $x^{a}_{RU, p,t}$ \\
   Emissions score of company $p$ in sector $a$ and year $t$ & $x^{a}_{EM, p,t}$ \\
   Environmental innovation score of company $p$ & $x^{a}_{EI, p,t}$ \\
 in sector $a$ and year $t$  & \\
   Workforce score of company $p$ in sector $a$ and year $t$ & $x^{a}_{WF, p,t}$ \\
   Human rights score of company $p$ in sector $a$ and year $t$ & $x^{a}_{HR, p,t}$ \\
   Community score of company $p$ in sector $a$ and year $t$ & $x^{a}_{CO, p,t}$ \\
   Product Responsibility score of company $p$  & $x^{a}_{PR, p,t}$ \\
 in sector $a$ and year $t$ & \\
   Management score of company $p$ in sector $a$ and year $t$ & $x^{a}_{MG, p,t}$ \\
   Shareholders score of company $p$ in sector $a$ and year $t$ & $x^{a}_{SH, p,t}$ \\
   CSR strategy score of company $p$ in sector $a$ and year $t$ & $x^{a}_{CS, p,t}$ \\
    Total number of zero values in ESG categories  & $x^{a}_{zero, p,t}$ \\
    of company $p$ in sector $a$ and year $t$ & \\
        Set of ESG category indices & $S_{CAT} = \{RU, EM, EI, WF, HR, CO, PR, MG, SH, CS\}$ \\
    Set of total number of zero values in ESG categories  & $S^{a}_{zero,t} = \{x^{a}_{zero, 1,t}, \ldots, x^{a}_{zero, n_a,t}\}$ \\
of companies in sector $a$ and year $t$ & \\
Total number of companies that company $p$ has  & $e^{a}_{p,t}$\\
the same total number of zero &\\
 values in ESG categories in sector $a$ and year $t$ & \\
Total number of companies that company $p$ has   & $l^{a}_{p,t}$\\ 
a higher total number of zero&\\ 
values in ESG categories in sector $a$ and year $t$ & \\  \hline
\end{tabular}}
\caption{Mathematical indices, data, and their notation used in the paper.}
\label{table6}
\end{table}

\textit{Total number of zero values in ESG categories of company $p$ in sector $a$ and year $t$ ($x^{a}_{zero, p,t}$):}
\begin{equation}
x^{a}_{zero, p,t} =  \smashoperator{\sum_{\substack{j’ \in S_{CAT}\\j': x^{a}_{j', p,t}=0}}} 1,\quad \forall{a,p,t}. \
\label{eq:find-zero}
\end{equation}
\textit{Total number of companies that company $p$ has the same total number of zero values in ESG categories in sector $a$ and year $t$ ($e^{a}_{p,t}$):}
\begin{equation}
e^{a}_{p,t} =  \smashoperator{\sum_{\substack{j’ \in [1, n_a] \\j': x^{a}_{zero, j',t}=x^{a}_{zero, p,t}}}} 1,\quad \forall{a,p,t}. 
\label{eq:equal-zero}
\end{equation}
\textit{Total number of companies that company $p$ has a higher total number of zero values in ESG categories in sector $a$ and year $t$ ($l^{a}_{p,t}$):}
\begin{equation}
l^{a}_{p,t} =  \smashoperator{\sum_{\substack{j’ \in [1, n_a] \\j': x^{a}_{zero, j',t}<x^{a}_{zero, p,t}}}} 1, \quad \forall{a,p,t}. 
\label{eq:higher-zero}
\end{equation}

\section{95$\%$ VaR, volatility, \textit{vvrisk}}\label{app-ratio}

\begin{figure}[H]
     \caption{Pairwise scatter plots of volatility, VaR, and \textit{vvrisk} in a year in S\&P 500, where upper diagonal shows the empirical Kendall's $\tau$.}
\includegraphics[width=.3\textwidth]{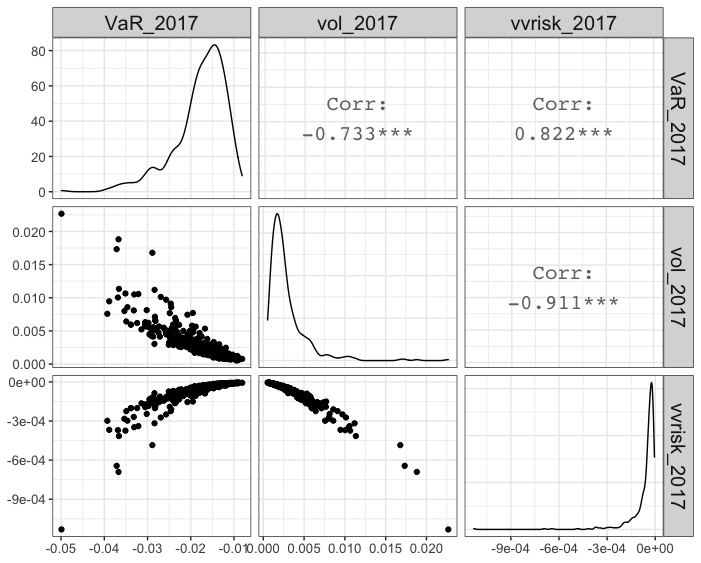} \hfill
\includegraphics[width=.3\textwidth]{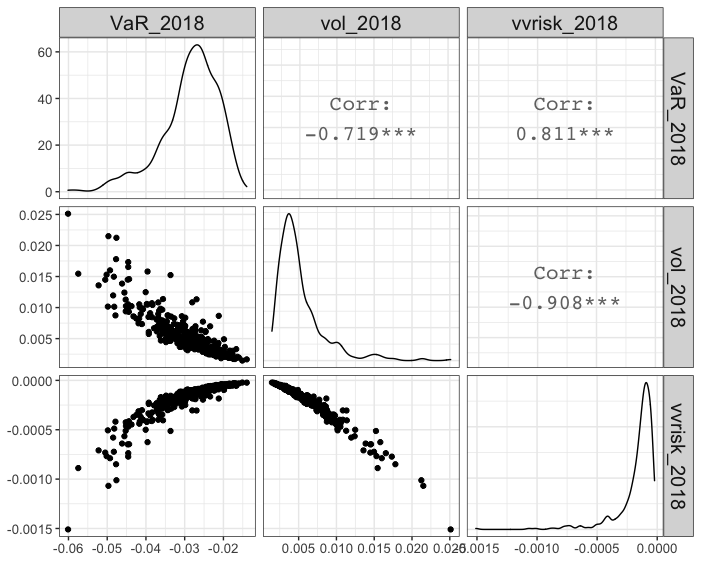}  \hfill
\includegraphics[width=.3\textwidth]{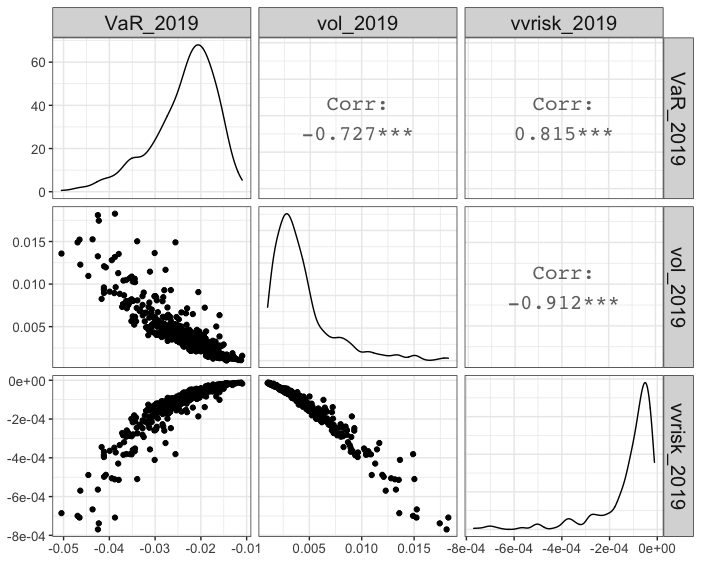} 
\label{fig:esg-ratio}
\end{figure}

\begin{figure}[H]
\centering
     \caption{95\% VaR across ten sectors in S\&P 500 (top) and EuroStoxx 600 (bottom).}
\includegraphics[width=\textwidth]{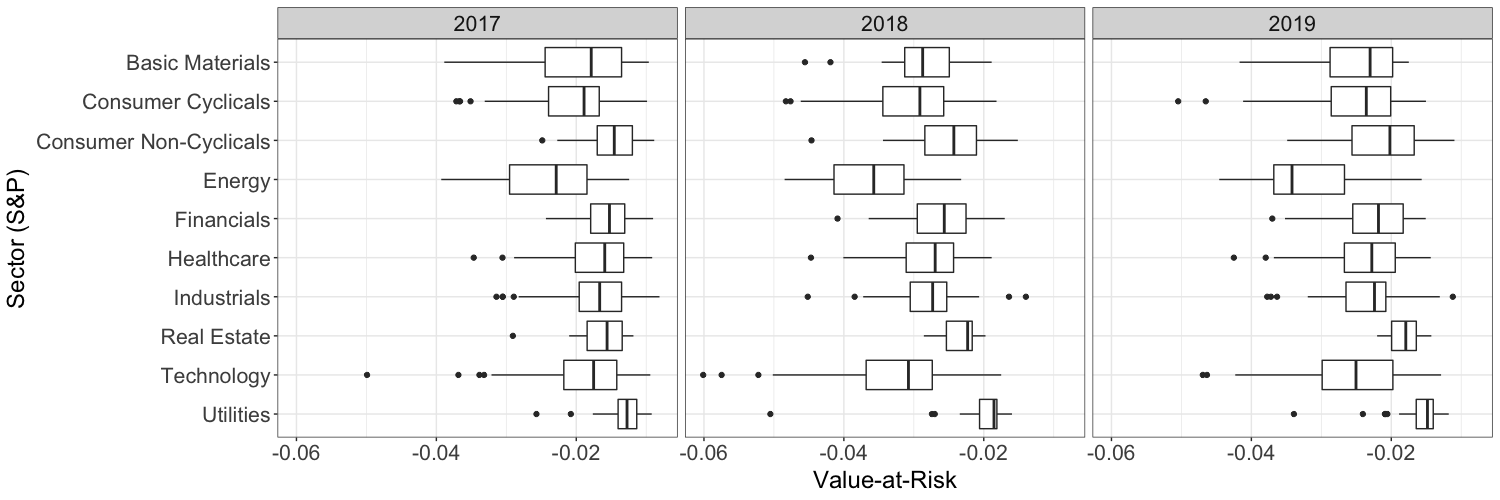}
\includegraphics[width=\textwidth]{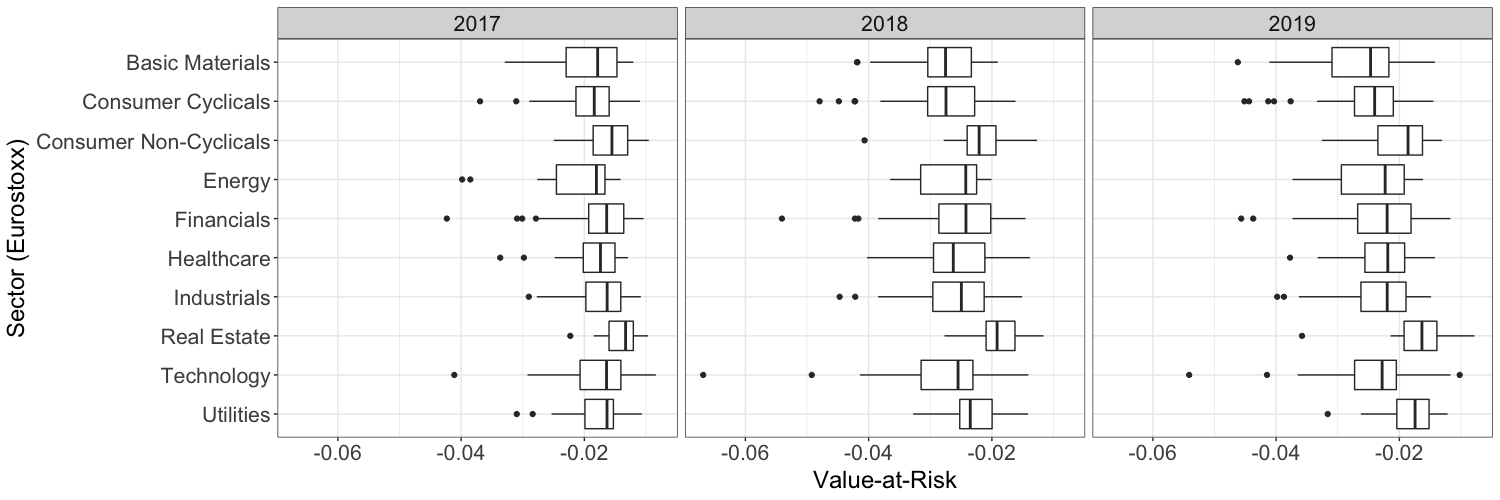}
   \label{fig:data-VaR}
\end{figure}   

\section{Proof of the M pillar score bounds and average}\label{app-proof}
Let $S=\{x_1, \ldots, x_N\}$ be a set with $N$ elements such that $x_1 \leq \ldots \leq x_N$. For the $d$th element, $x_d$, assume the number of elements whose value is smaller/larger than $x_d$ in $S$ are denoted by $n^{d}_s$/$n^{d}_l$. Also, $n^{d}_e$ corresponds to the number of elements $x_d$ has the same value  in $S$ (including itself). It holds that $n^{d}_s + n^{d}_l + n^{d}_e = N$ for $d=1, \ldots N$. Then its M pillar score is given by:
\begin{align*}
x^{d}_{M} = 100 \cdot \frac{n^{d}_s + \frac{n^{d}_e}{2}}{N} \quad \textrm{for} \quad  d =1, \ldots, N.
\end{align*}
\textit{Lower bound on M pillar score:} Since it holds $n^{d}_s \geq 0, n^{d}_e  \geq 0, N  \geq 0$, we have $x^{d}_{M}\geq 0$ for $\forall_d$, \\
\textit{Upper bound on M pillar score:} Since it holds $x^{d}_{M} = 100 \cdot \frac{n^{d}_s + \frac{n^{d}_e}{2}}{N} \leq 100 \cdot \frac{n^{d}_s + \frac{n^{d}_e}{2}}{n^{d}_s + n^{d}_e} \leq   100 \cdot \frac{n^{d}_s +n^{d}_e}{n^{d}_s + n^{d}_e}$, we have $x^{d}_{M}\leq 100$ for $\forall_d$. \\
\textit{Average value of M pillar score:} Denoting the average value of the M pillar score for the elements in $S$ by $\bar{x}_{M}$, we can write:
\begin{align*}
\bar{x}_{M} = 100 \cdot \frac{(n^{1}_s + \ldots + n^{N}_s)+ \frac{(n^{1}_e + \ldots + n^{N}_e)}{2}}{N \cdot N}.
\end{align*}
In the first scenario, assume $x_1 < \ldots < x_N$. Then, it holds
\begin{align*}
\bar{x}_{M} &= 100 \cdot \frac{(0 + \ldots + N-1)+ \frac{(1 + \ldots + 1)}{2}}{N \cdot N} = 100 \cdot \frac{\frac{N \cdot (N-1)}{2}+ \frac{N}{2}}{N \cdot N} = 50.
\end{align*}
In the second scenario, assume $x_1 =\ldots = x_j < x_{j+1} < \ldots <x_N$. Then, we have
\begin{align*}
\bar{x}_{M} &= 100 \cdot \frac{(0 + \ldots +0 + J + (J+1)+ \ldots + N-1)+ \frac{(J + \ldots +J + 1 + \ldots +1)}{2}}{N \cdot N} 
\\ &= 100 \cdot \frac{\frac{(N-1+J) \cdot (N-J)}{2}+ \frac{J \cdot J + (N-J)}{2}}{N \cdot N} = 50.
\end{align*}
The second scenario can be easily adopted for the equal elements, which exist more than once in the set, and it can be proven that the average M pillar score is 50.

\section{Missing values in ESG categories}\label{app-missing-esg}
\begin{table}[H]
\centering
\scalebox{0.7}{
\begin{tabular}{lcccccc}
 \hline
 $t = 2017$& Resource & Emissions & Environmental  & Human& Product  & CSR \\ 
 S\&P (EuroStoxx) & Use &  & Innovation &   Rights& Responsibility &   Strategy \\ 
  \hline
Basic Materials & 0\% (0\%) & 4\% (2\%) & 26\% (18\%) & 9\% (18\%) & 4\% (2\%) & 4\% (2\%) \\ 
  Consumer Cycl. & 16\% (4\%) & 13\% (5\%) & 43\% (29\%) & 27\% (28\%) & 0\% (10\%) & 34\% (6\%) \\ 
  Consumer N-Cycl. & 0\% (0\%) & 3\% (0\%) & 15\% (20\%) & 8\% (24\%) & 0\% (0\%) & 8\% (2\%) \\ 
  Energy & 4\% (0\%) & 0\% (0\%) & 62\% (29\%) & 38\% (12\%) & 8\% (0\%) & 4\% (0\%) \\ 
  Financials & 18\% (9\%) & 18\% (3\%) & 42\% (28\%) & 52\% (35\%) & 0\% (11\%) & 30\% (3\%) \\ 
  Healthcare & 14\% (0\%) & 20\% (0\%) & 66\% (55\%) & 29\% (36\%) & 0\% (6\%) & 29\% (3\%) \\ 
  Industrials & 14\% (4\%) & 12\% (2\%) & 32\% (21\%) & 28\% (33\%) & 3\% (6\%) & 25\% (8\%) \\ 
  Real Estate & 18\% (4\%) & 14\% (4\%) & 25\% (23\%) & 61\% (69\%) & 4\% (35\%) & 29\% (15\%) \\ 
  Technology & 12\% (2\%) & 21\% (2\%) & 29\% (27\%) & 18\% (22\%) & 0\% (9\%) & 40\% (7\%) \\ 
  Utilities & 3\% (0\%) & 3\% (0\%) & 14\% (4\%) & 34\% (32\%) & 0\% (0\%) & 7\% (0\%) \\     \hline

 $t = 2018$& Resource & Emissions & Environmental  & Human& Product  & CSR \\ 
  S\&P (EuroStoxx) & Use &  & Innovation &   Rights& Responsibility &   Strategy \\ 
  \hline
Basic Materials & 0\% (0\%) & 4\% (0\%) & 26\% (18\%) & 4\% (8\%) & 0\% (0\%) & 4\% (2\%) \\ 
   Consumer Cycl. & 16\% (4\%) & 10\% (3\%) & 44\% (29\%) & 21\% (15\%) & 0\% (4\%) & 30\% (3\%) \\ 
 Consumer N-Cycl.& 0\% (0\%) & 0\% (0\%) & 15\% (17\%) & 3\% (20\%) & 0\% (0\%) & 5\% (0\%) \\ 
  Energy & 0\% (0\%) & 0\% (0\%) & 62\% (29\%) & 25\% (12\%) & 4\% (0\%) & 4\% (0\%) \\ 
  Financials & 13\% (7\%) & 15\% (2\%) & 42\% (19\%) & 42\% (22\%) & 0\% (6\%) & 22\% (3\%) \\ 
  Healthcare & 7\% (0\%) & 12\% (0\%) & 62\% (48\%) & 21\% (18\%) & 0\% (0\%) & 25\% (0\%) \\ 
  Industrials & 12\% (4\%) & 9\% (2\%) & 34\% (20\%) & 18\% (21\%) & 2\% (6\%) & 28\% (5\%) \\ 
  Real Estate & 11\% (4\%) & 11\% (0\%) & 25\% (19\%) & 43\% (58\%) & 0\% (31\%) & 11\% (0\%) \\ 
  Technology & 9\% (2\%) & 20\% (0\%) & 29\% (29\%) & 17\% (16\%) & 0\% (9\%) & 34\% (4\%) \\ 
  Utilities & 0\% (0\%) & 3\% (0\%) & 14\% (11\%) & 31\% (11\%) & 0\% (0\%) & 3\% (0\%) \\    \hline

 $t = 2019$& Resource & Emissions & Environmental  & Human& Product  & CSR \\ 
 S\&P (EuroStoxx)  & Use &  & Innovation &   Rights& Responsibility &   Strategy \\ 
  \hline
Basic Materials & 0\% (0\%) & 4\% (0\%) & 26\% (18\%) & 4\% (2\%) & 0\% (0\%) & 0\% (2\%) \\ 
   Consumer Cycl. & 14\% (4\%) & 8\% (1\%) & 44\% (27\%) & 16\% (10\%) & 0\% (1\%) & 26\% (0\%) \\ 
  Consumer N-Cycl. & 0\% (0\%) & 0\% (0\%) & 13\% (17\%) & 0\% (5\%) & 0\% (0\%) & 5\% (0\%) \\ 
  Energy & 0\% (0\%) & 0\% (0\%) & 54\% (24\%) & 21\% (6\%) & 4\% (0\%) & 0\% (0\%) \\ 
  Financials & 8\% (5\%) & 8\% (1\%) & 35\% (16\%) & 32\% (15\%) & 0\% (1\%) & 10\% (2\%) \\ 
  Healthcare & 4\% (0\%) & 4\% (0\%) & 59\% (45\%) & 11\% (3\%) & 0\% (0\%) & 14\% (0\%) \\ 
  Industrials & 8\% (1\%) & 5\% (1\%) & 31\% (19\%) & 14\% (11\%) & 2\% (4\%) & 23\% (2\%) \\ 
  Real Estate & 11\% (4\%) & 4\% (0\%) & 18\% (15\%) & 43\% (42\%) & 0\% (19\%) & 7\% (0\%) \\ 
  Technology & 7\% (2\%) & 15\% (0\%) & 26\% (27\%) & 12\% (11\%) & 0\% (7\%) & 20\% (2\%) \\ 
  Utilities & 0\% (0\%) & 3\% (0\%) & 10\% (7\%) & 24\% (4\%) & 0\% (0\%) & 3\% (0\%) \\    \hline
\end{tabular}}
\caption{Percentage of the companies whose ESG category score is missing (zero) classified by year, sector and category in S\&P 500 and EuroStoxx 600. Only the categories for which at least a company in a given year does not contain any information are reported.}
\label{table7}
\end{table} 

\section{Results for the sectors with the zero M pillar weight}\label{app-m-zero}
\begin{table}[H]
\centering
\setlength{\tabcolsep}{0.5pt}
\begin{tabular}{lcccc}
  \hline
Sector (S\&P)  & E & S & G & M \\ 
  \hline
Basic Materials & 0.800 & 0.100 & 0.100 & 0.000 \\ 
  Consumer N-Cycl.& 0.486 & 0.410 & 0.104 & 0.000 \\ 
  Financials & 0.100 & 0.100 & 0.800 & 0.000 \\ 
  Healthcare & 0.730 & 0.100 & 0.169 & 0.000 \\ 
  Real Estate & 0.100 & 0.800 & 0.100 & 0.000 \\ 
   \hline
\end{tabular}
\quad
\setlength{\tabcolsep}{1.5pt}
\begin{tabular}{lcccc}
  \hline
Sector (EuroStoxx)  & E & S & G & M \\ 
  \hline
Basic Materials & 0.100 & 0.618 & 0.282 & 0.000 \\ 
  Consumer N-Cycl. & 0.683 & 0.217 & 0.100 & 0.000 \\ 
 Technology & 0.790 & 0.100 & 0.110 & 0.000 \\ 
  Utilities & 0.233 & 0.542 & 0.225 & 0.000 \\ 
               &  &     &     &  \\
   \hline
\end{tabular}
 \caption{New E, S, G, M pillar score weights, resulting in the ESGM scores in Tables 9 and 10 across sectors for which we have a zero M pillar score weight in S\&P 500 (left) and EuroStoxx 600 (right).}
\label{table8}
\end{table}

\begin{table}[H]
\centering
    \begin{tabular}{lrlrl}
    \hline
     & \multicolumn{4}{c}{Panel A: ESG and VaR} \\
      \hline
Sector (S\&P) &2017 &&2018   \\
\hline
Basic Materials    &     0.130&& 0.099& \\
Consumer N-Cycl.&-0.047 && 0.001& \\
Financials      &        0.020 &&-0.160& \\
Healthcare   &           0.309 &***& 0.278&*** \\
Real Estate     &       -0.138 &&-0.312& \\
\hline
Sector (EuroStoxx) &2017 &&2018   \\
\hline
Basic Materials    &    -0.096&& -0.171& \\
Consumer N-Cycl.   &0.054 && 0.085&  \\
Technology      &        0.089 && 0.168&* \\
Utilities           &    0.127 && 0.296&** \\
\hline
  \end{tabular}
      \begin{tabular}{rlrl}
      \hline
      \multicolumn{4}{c}{Panel B: ESG(M) and VaR} \\
      \hline
2017 &&2018   \\
\hline
       0.296&**& 0.194& \\
0.015 &&0.053& \\
           0.009 &&-0.121& \\
           0.297 &***& 0.282&*** \\
          -0.032 &&-0.286& \\
\hline
2017 &&2018   \\
\hline
        -0.056&& -0.099& \\
        0.129&& 0.115& \\
           0.117 && 0.230& **\\
                     0.206&*&  0.333&*** \\
           \hline
  \end{tabular}
    \caption{Kendall's $\tau$ between ESG, ESGM scores, and 95\% VaR in 2017, 2018 across sectors (in-sample) for which we have a zero M pillar score weight in S\&P 500 (top) and EuroStoxx 600 (bottom). The hypothesis testing is $H_0:\tau=0$ versus $H_A:\tau>0$. *, **, and *** denote statistical significance at 10\%, 5\%, and 1\% levels, respectively.}
    \label{table9}
\end{table}

\begin{table}[H]
\centering
    \begin{tabular}{lrlrl}
    \hline
     & \multicolumn{3}{c}{Panel A: ESG and VaR} \\
      \hline
Sector (S\&P500) &2019 &  \\
\hline
Basic Materials    &    -0.075 & \\
Consumer N-Cycl. & 0.115    & \\
Financials   &   -0.102  & \\
Healthcare    &    0.161 & **\\
Real Estate    &    -0.310 & \\
\hline
Sector (EuroStoxx) &2019 &  \\
\hline
Basic Materials   &      -0.064&  \\
Consumer N-Cycl. & 0.168&*  \\
Technology    &        0.073&  \\
Utilities       &         0.270&  **\\
\hline
  \end{tabular}
      \begin{tabular}{rlrl}
      \hline
      \multicolumn{2}{c}{Panel B: ESG(M) and VaR} \\
      \hline
2019 &   \\
\hline
0.043&   \\
0.115&   \\
-0.031&   \\
0.088&   \\
-0.259&   \\
\hline
2019 &   \\
\hline
-0.040&  \\
 0.200& ** \\
0.101 &  \\
0.286&  **\\
\hline
  \end{tabular}
      \caption{Kendall's $\tau$ between ESG, ESGM scores, and 95\% VaR in 2019 across sectors (out-of-sample) for which we have a zero M pillar score weight in S\&P 500 (top) and EuroStoxx 600 (bottom).} 
          \label{table10}
\end{table}
\end{document}